\documentclass[lettersize,journal]{IEEEtran}
\usepackage{textcomp}
\usepackage{amsmath,amssymb}

\usepackage{amsthm}
\usepackage{amsfonts,mathrsfs}
\usepackage{color}
\usepackage{graphicx}
\usepackage{epsfig}
\usepackage{subcaption}
\usepackage{enumerate}
\usepackage{cancel}

\let\labelindent\relax
\usepackage{enumitem}
\setlist{leftmargin=1.5mm} 
\setlist{nolistsep} 

\usepackage{algorithm,algorithmicx}
\usepackage{algpseudocode}
\usepackage[hidelinks]{hyperref}
\usepackage{marginnote}
\usepackage{empheq}
\usepackage{bm}
\usepackage{accents}
\usepackage{cite}
\usepackage{balance}
\usepackage{latexsym}
\usepackage{graphicx}
\usepackage{float}
\usepackage{colortbl}
\usepackage{hyperref}
\usepackage{enumerate}
\usepackage{multicol}
\usepackage{float}
\usepackage{cite}
\usepackage{cleveref}
\usepackage[normalem]{ulem} 

\usepackage[usenames,dvipsnames,svgnames,table]{xcolor}
\usepackage{tikz}
\usepackage{pgfplots}
\pgfplotsset{compat=newest} 
\pgfplotsset{plot coordinates/math parser=false} 
\usepgfplotslibrary{patchplots}
\usetikzlibrary{automata,positioning}

\newcommand{\citep}{\cite}

\DeclareMathOperator{\proj}{proj}

\usepackage{soul}

\newcommand{\bs}{\boldsymbol}
\newcommand{\mc}{\mathcal}
\newcommand{\bb}{\mathbb}
\newcommand{\R}{\bb R}
\newcommand{\N}{\mathbb{N}}
\newcommand{\E}{\mathbb{E}}

\DeclareMathAlphabet{\mathbbmsl}{U}{bbm}{m}{sl}

\newcommand{\diag}{\operatorname{diag}}

\newcommand{\col}{\operatorname{col}}
\newcommand{\zer}{\operatorname{zer}}

\newcommand{\nc}{\mathrm{N}}

\newcommand{\Rmnum}[1]{\expandafter\@slowromancap\romannumeral #1@}

\newcommand{\qedd}{\ensuremath{\hfill \blacksquare}}

\newcommand{\sigmaM}{\sigma^{\text{M}}}
\newcommand{\sigmar}{\sigma^{\rho }  }
\newcommand{\bOmega}{\boldsymbol{\Omega}}
\newcommand{\nx}{n_{\text{x}}}
\renewcommand{\nu}{n_{\text{u}}}
\newcommand{\Se}{S_{\mathrm{edge}}}

\newcommand{\rhoeq}{\rho_{\mathrm{eq}}} 
\newcommand{\ueq}{u_{\mathrm{eq}}}

\newcommand{\uerr}{\tilde{u}}

\newcommand{\potential}{p}
\renewcommand{\tilde}{\widetilde}

\newcommand{\bx}{\boldsymbol{x}}
\newcommand{\by}{\boldsymbol{y}}

\newcommand{\bu}{\boldsymbol{u}}

\newcommand{\tausp}[1]{\tau^{\mathrm{kp}}_{#1}}
\newcommand{\bstausp}{\bs{\tau}^{\mathrm{kp}}}
\newcommand{\musp}[1]{\mu^{\mathrm{kp}}_{#1}}
\newcommand{\bmusp}{\bs{\mu}^{\mathrm{kp}}}
\newcommand{\NP}[1]{\mathrm{KP}_{#1}}
\newcommand{\PN}[1]{\mathrm{PK}_{#1}}
\newcommand{\usp}[1]{{u}^{\mathrm{kp}}_{#1}}
\newcommand{\barusp}{\bs{u}^{\mathrm{kp}}}

\newcommand{\bomega}{\bm{\omega}}

\renewcommand{\P}{\mathbb{P}}

\newcommand{\ones}{\bs 1}

\usepackage{changepage}

\makeatletter
\newcommand{\specialcell}[1]{\ifmeasuring@#1\else\omit$\displaystyle#1$\ignorespaces\fi}
\makeatother
\newtheorem{assumption}{Assumption}
\newtheorem{proposition}{Proposition}

\newtheorem{definition}{Definition}
\newtheorem{lemma}{Lemma}

\newtheorem{theorem}{Theorem}{\it}{}

\newtheorem{remark}{Remark}

\makeatletter
\newcommand{\cbigoplus}{\DOTSB\cbigoplus@\slimits@}
\newcommand{\cbigoplus@}{\mathop{{\bigoplus}}}
\makeatother

\usepackage{setspace}
\usepackage{titlesec}
\titlespacing\section{0pt}{12pt plus 4pt minus 1pt}{0pt plus 2pt minus 2pt}
\titlespacing\subsection{0pt}{12pt plus 4pt minus 1pt}{0pt plus 2pt minus 2pt}
\titlespacing\subsubsection{0pt}{12pt plus 4pt minus 1pt}{0pt plus 2pt minus 2pt}

\allowdisplaybreaks
\title{Probabilistic Game-Theoretic Traffic Routing}
\author{Emilio Benenati, and Sergio Grammatico, \IEEEmembership{Senior Member, IEEE}
	\thanks{The authors are with the Delft Center of Systems and Control (DCSC), TU Delft, the Netherlands. E-mail addresses: \texttt{\{e.benenati, s.grammatico\}@tudelft.nl}. }
	\thanks{This work was partially supported by the ERC under research project COSMOS (802348). }
}

\begin{document}
	
	\maketitle
	\begin{abstract}
	We examine the routing problem for self-interested vehicles using stochastic decision strategies. By approximating the road latency functions and a non-linear variable transformation, we frame the problem as an aggregative game. We characterize the approximation error and we derive a new monotonicity condition for a broad category of games that encompasses the problem under consideration. Next, we propose a semi-decentralized algorithm to calculate the routing as a variational generalized Nash equilibrium and demonstrate the solution's benefits with numerical simulations. {In the particular case of potential games, which emerges for linear latency functions,} we explore a receding-horizon formulation of the routing problem, showing asymptotic convergence to destinations and analysing closed-loop performance dependence on horizon length through numerical simulations.
	\end{abstract}
\begin{IEEEkeywords}
Traffic control, Game theory, Variational methods
\end{IEEEkeywords}

	\section{Introduction}
	 \IEEEPARstart{T}{raffic} jams generate a heavy burden on the society \cite{barth_traffic_2009} and, as car traffic already makes up a large share of the EU transport infrastructure costs \cite{european_commission_and_directorate-general_for_mobility_and_transport_sustainable_2019}, it is imperative to mitigate the road congestion without expanding the existing infrastructure. The increased availability of real-time information on the state of the road network  has the potential for a more efficient traffic-aware route planning. \\
	 \emph{Related work:} {Previous studies have considered routing strategies which optimize a system-wide efficiency metrics \cite{jahn_optimal_1999, jahn_system-optimal_2005, angelelli_system_2021}. A shortcoming of this approach is that the drivers can find a more advantageous path than the one assigned and thus they might not adhere to such an externally-imposed solution. A workaround to this issue is to limit the inconvenience caused to the users \cite{jahn_system-optimal_2005, angelelli_system_2021}. However, this approach still does not fully address the inherently competitive nature of the problem, which is more properly modeled as a \emph{game}, as noted in the seminal work\cite{wardrop_theoretical_1952}.} Crucially, under relatively loose conditions, games admit  a set of Nash equilibria (or Wardrop equilibria, if the impact of  each agent on the road latency is assumed negligible), that is, a set of decision strategies from which no agent has an incentive in unilaterally deviating and thus need no external imposition. Wardrop equilibium-based routing methods have been studied first in \cite{wardrop_theoretical_1952}, and in \cite{roughgarden_how_2002-1} it is shown that they exhibit bounded system-level inefficiency. The inclusion of capacity constraints was considered in \cite{correa_selfish_2004}. {The Wardrop equilibrium of the routing problem is typically found by reformulating the problem as an equivalent optimization problem \cite{dafermos_traffic_1972} or variational inequality \cite{dafermos_traffic_1980}, and capacity constraints can be handled by Lagrangian duality \cite{larsson_augmented_1995}. However, these reformulations require a pre-computation of every route available between each origin-destination pair, which can become cumbersome for large networks.} {By contrast, \cite{calderone_markov_2017, li_tolling_2019} propose a Markov Decision Process (MDP) model that requires no pre-computation of the paths.} This idea is further elaborated in \cite{bakhshayesh_decentralized_2021},  where the problem is cast as a (monotone, aggregative) generalized Nash equilibrium problem (GNEP). {The latter approach is particularly appealing from a computational perspective, as recent developments in algorithmic game theory has made available a plethora of efficient Nash equilibrium-seeking algorithms which allow for a decentralized computation \cite{belgioioso_semi-decentralized_2023}. }\\
	 \emph{Contributions:} {Inspired by \cite{bakhshayesh_decentralized_2021}, we cast the vehicle routing problem as a GNEP. Our contribution is threefold:}
     \begin{itemize}[leftmargin=*]  \item
     	In Section \ref{sec:traffic_GNE}, we demonstrate that commonly used routing objective functions, e.g. \cite{calderone_markov_2017, bakhshayesh_decentralized_2021, li_tolling_2019} and this work, are an approximation of the expected traversing time and we characterize the approximation error.
    	\item In Section \ref{sec:GNE_seeking}, we establish the monotonicity of the game under a less restrictive condition than the one derived in \cite[Lemma 1]{bakhshayesh_decentralized_2021}. Technically, we achieve this by carefully characterizing the eigenvalues of a class of matrices whose structure emerges in the pseudogradient's Jacobian (Lemma \ref{le:eigenvalues_A}). We then solve the game via the Inertial Forward-Reflected-Backward (I-FoRB) algorithm \cite{belgioioso_semi-decentralized_2023}, which does not require the pseudo-gradient to be cocoercive as does the algorithm adopted in \cite{bakhshayesh_decentralized_2021}  (namely, the preconditioned forward-backward \cite{belgioioso_projected-gradient_2018}) and thus it converges without a quadratic regularization term \cite[Equation 5]{bakhshayesh_decentralized_2021}. 
    	\item  In Section \ref{sec:receding_horizon_GNE}, we propose a modified formulation of the problem which allows one to progressively recompute the agents' paths in a receding horizon fashion (instead of solving for the entire path in one computation) in the particular case of a  \emph{potential} game \cite{monderer_potential_1996}. This property of the game allows one to cast the receding horizon game as an MPC controller, and thus we show asymptotic convergence by a careful choice of the terminal cost. This novel approach allows one to reduce the decision horizon, thus reducing the computational burden as the vehicles move forward. 
    \end{itemize}

\noindent Finally, in Section \ref{sec:numerical}, we support the theoretical results by comparative numerical simulations.

\section{Notation}
 For a matrix $X$,we denote  its $(i,j)$-th element as $ [X]_{(i,j)} $ and its spectral norm as $\|\cdot\|$. The operators $ \col(x_i)_{i\in\mc I} $, $\mathrm{row}(x_i)_{i\in\mc I}$ denote the  column-wise and row-wise stack of $(x_i)_{i\in\mc I}$, respectively. We denote the block-diagonal matrix with blocks $(X_i)_{i\in\mc I}$  as $ \diag( X_i)_{i\in\mc I}$. We define $\mathrm{avg}((x_i)_{i\in\mc I}):= \frac{1}{|\mc I|}\sum_{i\in\mc I} x_i$. The vector in $\R^n$ with all elements $1$ ($0$) is denoted as $\ones_n$ ($\boldsymbol{0}_n$). The subscript is omitted when the dimension is clear. We denote the partial gradient of $f$ with respect to $x$ as $\nabla_x f$. If $f$ is scalar, its first derivative is $f'$. We denote the  Jacobian of $F$ as $DF$. The Cartesian product is denoted as $\times$ and the Minkowski sum as $+$.  \\
\emph{Operator theory:} Given $C\subset \R^n$, $N_C$ denotes its normal cone \cite[Def. 6.38]{bauschke_convex_2017}. The Euclidean projection onto $C$ is denoted by $\proj_C(x)$. For an operator $T:\R^n\rightrightarrows\R^n$, we denote $\zer(T):=\{x\in\R^n| \boldsymbol{0}_n \in T(x)\}$. The operator $T:C\to C$ is ($m$-strongly) monotone in $C$ if $\langle T(x) - T(y), x-y\rangle \geq m\| x-y\|^2$ for all $x,y\in C$, for some $m\geq 0$ $(>0)$. \\
\emph{Probability theory:} {Given a probability space $(\Omega, \mc F, \P)$ with sample space $\Omega$ and event set $\mc F$, let  $A, B\in\mc F$. Then, $\P[A]$  denotes the probability of $A$,  $\P[A|B]$  denotes the probability of $A$ conditioned on $B$, $\P[A,B]$ denotes the joint probability of $A$ and $B$ and $\E[X]$ denotes the expected value of $X:\Omega \to \R^n$, for some $n\in \mathbb N$. }We denote $\Delta^n:=\{x\in [0,1]^n: \ones_n^{\top}x =1\}$. 

	\section{Traffic routing as a Generalized Nash Equilibrium problem}
	\label{sec:traffic_GNE}
		Let $\mc R(\mc N, \mc E)$ a  directed graph modelling a road network whose  nodes $\mc N$ represent the junctions and each edge $(a,b)\in \mc E$ represents the road from $a$ to $b$.  We study the problem of routing $N$  populations of vehicles $\mc I:=\{1,...,N\}$. Denote $\mc I_{-i}:= \mc I \setminus \{i\}$ for all $i\in\mc I$. Each population is made up of $V$ vehicles, where vehicles in the same population $i\in\mc I$ share the same initial position $b_i \in \mc N$ and destination $d_i \in \mc N$. 
		\begin{remark}
			{Each population contains the same number of vehicles without loss of generality. In fact, let each population contain $(V_i)_{i\in\mc I}$ vehicles and let $V\in \mathbb{N}$ be such that $V_i/V\in \mathbb{N}$ for all $i$. Then, we can split each population $i$ into $V_i/V$ populations of equal size $V$.}
		\end{remark}
		Next, we ensure that each destination  node can be reached:
		\begin{assumption}
			\label{as:strongly_connected}
			$\mc R(\mc N, \mc E)$  is strongly connected and $(a,a)\in\mc E$ for each $a\in\mc N$.
		\end{assumption}
		The vehicles aim at reaching their destinations within a time horizon $T$. The control action determines the probability for the controlled vehicle to drive through a certain road and it is the same for each vehicle in a population. In this {setting,} each population acts as a single entity, thus, we refer  to each of them as an \emph{agent}. We stress that the route of each vehicle is a realization of the probabilistic control action, thus vehicles represented by the same agent might take different routes.  To formalize this, let us denote the junction visited by the $v$-th vehicle of agent $i$ at time $t$ as $s^{i,v}_t$, which is a stochastic variable {with event space $\mc N$ and probability vector} $\rho^i_t \in \Delta^{|\mc N|}$, that is, $[\rho^i_t]_a:=\P[ s^i_t=a]$ for any $a\in \mc N$. The control actions are the column-stochastic matrices $\Pi^i_t\in\R^{|\mc N| \times |\mc N|}$,  defined  as
	$$ [\Pi^i_t]_{(b,a)} = \P[s^{i,v}_{t+1} = b | s^{i,v}_t =a] ~~~~ \text{for all } a,b \in \mc N. $$
	From the law of total probability,
	 \begin{equation} 
	 	\label{eq:nonlin_dyn} \rho^i_{t+1} = \Pi^i_t \rho^i_t ~~~\text{for all}~ i\in\mc I.
	  \end{equation} 
 The initial state of agent $i$ is $\rho^i_1$, with only non-zero element $[\rho^i_1]_{b_i}=1$. 
In the remainder of this section we show that, under an appropriate reformulation of \eqref{eq:nonlin_dyn}, the problem that arises in the proposed {setting} can be cast as a GNEP.
  \subsection{Affine formulation of the system dynamics}
  \label{sec:dyn_reformulation}
  Similarly to the approach in \cite{benenati_tractable_2019}, we reformulate the nonlinear dynamics in \eqref{eq:nonlin_dyn} in terms of the transformed variables
 \begin{equation}
 	\label{eq:nonl_transform}
 	M^i_{t,(a,b)} := [\Pi^i_t]_{(b,a)} [\rho^i_t]_a
 \end{equation} 
defined for all $i\in\mc I, (a,b)\in\mc E, t\in\mc T:=\{1,...,T\}$. By the definition of conditional probability, we have
\begin{equation}
	\label{eq:joint_probability}
	M^i_{t, (a,b)} =  \P[s^{i,v}_{t+1} = b , s^{i,v}_t =a].
\end{equation}
In words,  $M^i_{t,(a,b)}$ represents the probability that, at time $t$, agent $i$ traverses the road from $a$ to $b$. Denoting $\mc T^+:= \mc T \cup  \{T+1\}$, the decision variables of each agent are:
\begin{align}
	\label{eq:def_x}
	\begin{split}
 \omega_i &:=  \begin{bmatrix}
 		\mathrm{col}(M^i_{t, (a,b)})_{\substack{(a,b)\in\mc E ,  t\in\mc T }}  \\
	\mathrm{col}\left(\rho^i_t\right)_{t\in \mc T^+} 
\end{bmatrix}.
\end{split}
\end{align}
Without loss of generality, $\omega_i$ in \eqref{eq:def_x} does not include any variable corresponding to $[\Pi^i_{t}]_{(b,a)}$ with $(a,b)\notin \mc E$, since the probability of traversing a non-existing road is zero. We denote in boldface the concatenation over $\mc I$ and with boldface and indexing $-i$  the concatenation over $\mc{I}_{-i}$, e.g. $ \bomega := \mathrm{col}(\omega_i)_{i\in\mc I}$, $\bomega_{-i} := \mathrm{col}(\omega_j)_{j\in \mc I_{-i}}$. We also define $n_{\omega}:= T|\mc E| + (T+1)|\mc N|$.
 The following lemma states that, by imposing appropriate linear constraints on $\bomega$, the transformation in  \eqref{eq:nonl_transform} can be inverted and the resulting  matrices $\Pi^i_t$  are coherent with the dynamics in \eqref{eq:nonlin_dyn}. {All proofs are provided in the Appendix.}
\begin{lemma}
	\label{le:M_reformulation}
	Let $\omega_i$ in \eqref{eq:def_x} satisfy:
	\begin{subequations}
		 \label{eq:M_constr}
    	\begin{align}
	    &	\textstyle\sum_{a:(a,b)\in\mc E} M^i_{t,(a,b)}  = [\rho^i_{t+1}]_b \hspace{-14pt}& \text{for all}~b\hspace{-1pt}\in\hspace{-1pt}\mc N, ~t\hspace{-1pt}\in\hspace{-1pt}\mc T; \label{eq:M_constr_1}\\
	    &	\textstyle\sum_{b:(a,b)\in\mc E} M^i_{t,(a,b)}  = [\rho^i_{t}]_a  \hspace{-14pt}&\text{for all}~a\hspace{-1pt}\in\hspace{-1pt}\mc N, ~t\hspace{-1pt}\in\hspace{-1pt}\mc T;\label{eq:M_constr_2}\\
	    &	M^i_{t,(a,b)} \geq 0 \hspace{-14pt}& \text{for all}~(a,b) \hspace{-1pt}\in\hspace{-1pt}\mc E, ~t\hspace{-1pt}\in\hspace{-1pt}\mc T;\label{eq:M_constr_3} \\
    	&\rho^i_1 \in \Delta^{|\mc N|},~[\rho^i_1]_{b_i}=1. \label{eq:M_constr_4} 
	    \end{align}
	\end{subequations}
Then,  {$\omega_i\in (\Delta^{|\mc E|})^T \times (\Delta^{|\mc N|})^{(T+1)}$} and {a choice of  $(\Pi^i_t)_{i\in\mc I, t\in \mc T}$ such that $\rho_t^i$ follows the dynamics in \eqref{eq:nonlin_dyn} is:}
\begin{align}
	\label{eq:inverse_transform}
	&[\Pi^i_t]_{(b,a)} = \begin{cases}
		\frac{1}{|\mc N|} & \text{if}~[\rho^i_t]_a = 0 \\
		\frac{M^i_{t, (a,b)}}{[\rho^i_t]_a} & \text{if}~[\rho^i_t]_a \neq 0
	\end{cases}  
\end{align}  
for all ${ (a, b) \in\mc E, t \in \mc T,  i\in\mc I}$.
\end{lemma}
{Note that, in \eqref{eq:inverse_transform}, the $a$-th column of $\Pi_t^i$  such that $[\rho_t^i]_a=0$ can be chosen to be anything that sums to $1$, as those values do not influence the evolution of the vehicle distribution.}

\subsection{Control objective and constraints}
\label{sec:objective_and_constraints}
	We enforce the routing of each agent by constraining the destination to be reached with high probability:
	\begin{equation}
		\label{eq:final_dest_constr}
		\left[\rho^i_{T+1}\right]_{d_i} \geq 1-\varepsilon,
	\end{equation}
	where $\varepsilon$ is a free design parameter. Let us introduce the latency function $\ell_{(a,b)} :\R_{\geq 0}\to \R_{\geq 0}$, which maps the ratio of vehicles on a road to its traversing time. A common model is the Bureau of Public Transport (BPT) function \cite{u_s_b_of_public_roads_traffic_1964}:
	\begin{equation}
		\label{eq:bpt_latency}
		\ell_{(a,b)}^{\text{BPT}}(\sigma) := \tau_{(a,b)} \left ( 1 + 0.15 \left ( \tfrac{ \sigma + \zeta_{(a,b)}}{c_{(a,b)}} \right )^{\xi+1} \right) ,
	\end{equation}
	where $c_{(a,b)}$ and $\tau_{(a,b)}$  are the capacity and the free-flow traversing time of $(a,b)$, respectively, $\zeta_{(a,b)}\geq 0$ is the number of uncontrolled vehicles on the road normalized by  $VN$ and $\xi\geq 0$ is a parameter often set to $\xi=3$, e.g. \cite{u_s_b_of_public_roads_traffic_1964, angelelli_system_2021}.
	 More generally, we consider functions that satisfy the following:
	\begin{assumption} \label{as:ell}
		 For each $(a,b)\in\mc E$, {the latency function} $\ell_{(a,b)}$ is $C^2$, non-negative, non-decreasing and convex. 
	\end{assumption}
The number of vehicles traversing road $(a,b)$ at time $t$ is drawn from a Poisson's binomial distribution with $NV$ trials, grouped into $N$ groups of $V$ trials with identical success probability $M_{t,(a,b)}^i$, $i\in\{1,...,N\}$. Thus, its expected value is $\sum_{i\in\mc I} VM^i_{t, (a,b)}$ \cite[Eq. 15]{wang_number_1993} and the expected ratio of vehicles on $(a,b)$ is $\tfrac{\sum_{i} M^i_{t, (a,b)}}{N}=: \sigmaM_{(a,b), t}$.
Let $\hat{\sigma}^{\text{M}}_{t,(a,b)}$ the (actual) ratio of vehicles on road $(a,b)$ at time $t$. The expected traversing time is $\E [\ell_{(a,b)}(\hat{\sigma}^{\text{M}}_{(a,b), t})]$. This is in general intractable to compute: let us instead consider the first-order approximation of $\ell_{(a,b)}$ around the expected value of the argument.
 \begin{align}
 	\label{eq:first_order_approx}
 	\begin{split}
	&\E [\ell_{(a,b)}(\hat{\sigma}^{\text{M}}_{(a,b), t})] \simeq  \E\left[ \ell_{(a,b)}( \sigmaM_{(a,b),t}  ) + \right. \\
	& \left. \nabla \ell_{(a,b)}(\sigmaM_{(a,b),t}  ) (\hat{\sigma}^{\text{M}}_{(a,b), t} - \sigmaM_{(a,b),t}  ) \right] \overset{\{1\}}{=} \\
	& \ell_{(a,b)}(\sigmaM_{(a,b),t}) \hspace{-2pt}+ \hspace{-2pt} \nabla \ell_{(a,b)}(\sigmaM_{(a,b),t})(\E[\hat{\sigma}^{\text{M}}_{(a,b), t}]  \\ 
	& - \sigmaM_{(a,b),t}  ) \overset{\{2\}}{=} \ell_{(a,b)}(\sigmaM_{(a,b),t})
	\end{split}
 \end{align}
where in \eqref{eq:first_order_approx}, $\{1\}$  follows from the linearity of the expected value and from the fact that $\sigmaM_{(a,b), t}$ is deterministic, while $\{2\}$ follows from $\E[\hat{\sigma}_{(a,b),t}^{\text{M}}]=\sigma^{\text{M}}_{(a,b),t}$. Although nonlinear functions of the congestion were previously used as road traversing cost \cite{calderone_markov_2017, li_tolling_2019, bakhshayesh_decentralized_2021}, the interpretation {provided by \eqref{eq:first_order_approx}} {is novel, to the best of our knowledge.} To justify the approximation in \eqref{eq:first_order_approx}, we leverage known results on the Taylor series of stochastic functions \cite[\S 6]{wolter_introduction_2007} in order to show that the {approximation error} vanishes with the number of vehicles $NV$:
\begin{proposition} 
	\label{prop:justification_approx}
	{Let each $v$ in agent $i$ draw the event $(s_{t+1}^{i,v}=b, s_{t}^{i,v}=a)$ with probability $M_{t,(a,b)}^i$ independently from the remaining vehicles. Then, $(\ell_{(a,b)}(\sigmaM_{(a,b), t})-\ell_{(a,b)}(\hat{\sigma}^M_{(a,b), t}))^2 = y_{NV} + z_{NV} $, where $\E[y_{NV}] \leq \frac{1}{4NV}\nabla \ell(\sigmaM_{(a,b), t})^2 $ and, for every $\epsilon>0$, there exists $K_{\epsilon}>0$ such that 
		\begin{equation*}
		\mathrm{sup}_{NV\in\N}\left( \P \left[|z_{NV}|\geq \tfrac{K_{\varepsilon}}{8(NV)^{3/2}}\right]\right) \leq \epsilon.
	\end{equation*}  }
\end{proposition}
We now define the cost of traversing $(a,b)$ at time $t$:
 \begin{align}
	\label{eq:def_scalar_cost}
	\begin{split}
		J_{(a,b)}(M^i_{t,(a,b)},  \bs M^{-i}_{t, (a,b)} ):=
		& M^i_{t, (a,b)} \ell_{(a,b)}( \sigmaM_{(a,b),t}).
	\end{split}
\end{align}
The objective pursued by each agent reads then as follows:
\begin{equation}
	\label{eq:obj_func}
	J_i:= f_i(\omega_i) + \textstyle\sum_{\substack{(a,b)\in\mc E, t\in\mc T}} J_{(a,b)}(M^i_{t, (a,b)},  \bs M^{-i}_{t,(a,b)} ),
\end{equation}
where $f_i: \R^{n_{\omega}}\to \R$ encodes a local cost for agent $i$. Quadratic costs are considered in \cite[Eq. 5]{bakhshayesh_decentralized_2021}. We consider a more general class of functions.
\begin{assumption}
	\label{as:local_cost}
	The functions $(f_i)_{i\in\mc I}$ in \eqref{eq:obj_func} are convex and $C^2$.
\end{assumption}

Finally, we introduce the maximum capacity constraints
\begin{equation}
	\label{eq:shared_constr}
		\textstyle\sum_{i\in\mc I} M^i_{t, (a,b)} \leq \bar{c}_{(a,b)} ~~~ \text{for all } t\in\mc T, (a,b)\in \mc E
\end{equation}
which we recast via appropriately defined matrices  $(A_i)_{i\in\mc I}$, $A_i \in \R^{T|\mc E|\times n_{\omega}} $, $b\in \R^{T|\mc E|}$, $A:= \mathrm{row}(A_i)_{i\in\mc I}$:
\begin{equation}
	 \textstyle\sum_{i\in\mc I} A_i \omega_i = A\bomega \leq b. 
\end{equation}
\subsection{ Generalized Nash equilibrium problem }
Formalizing the model derived in Sections \ref{sec:dyn_reformulation} and \ref{sec:objective_and_constraints}, each agent solves the local optimization problem
\begin{subequations}
	\label{eq:generalized_game}
	\begin{empheq} [left={{\forall i \in \mc I} \colon \empheqlbrace\,}]{align}
		\underset{\omega_i \in \Omega_{i}}{\min} \quad & J_{i}(\omega_i, \bomega_{-i})  \label{eq:generalized_game:cost} \\
		\operatorname{s.t.} \quad & A_i\omega_i\leq b-\textstyle\sum_{j\in\mc I_{-i}} A_j \omega_j, \label{eq:generalized_game:constraint}
	\end{empheq}
\end{subequations}
where $\Omega_i := \left\{ \omega\in \R^{n_{\omega}} | \eqref{eq:M_constr}, \eqref{eq:final_dest_constr} ~\text{hold}  \right\}$ for all $i$.
The coupling between the   $N$  optimizations problems in \eqref{eq:generalized_game} emerges both in the cost functions and in the constraints, thus defining a \emph{generalized game} \cite{facchinei_generalized_2010}. In particular, the game is \emph{aggregative} because the coupling between cost functions depends only on the average decision $\sigmaM_{(a,b),t}$ for all $(a,b)$ and $t$. A desirable solution is the generalized Nash equilibrium  (GNE) $\bomega^*$, from which no agent has an incentive to unilaterally deviate. 
 \begin{definition}
 	\label{def:gne}
 	A collective strategy $\bomega^*\in\bOmega:= \left(\bigtimes_{i\in\mc I} \Omega_i \right) \bigcap \{ \bomega\in\R^{Nn_{\omega}} | A\bomega \leq b \} $ is a Nash equilibrium for the game in \eqref{eq:generalized_game} if, for each $ i\in\mc I$,
 	\begin{equation*}
 		J_i\left(\omega_i^*, \bomega^*_{-i} \right) \leq J_i\left(\omega_i,  \bomega_{-i}^* \right)
 	\end{equation*}
 	for any $\omega_i\in \Omega_i \bigcap \{y\in \R^{n_{\omega}}| A_iy\leq b-\sum_{j\in\mc I_{-i}} A_j \omega_j^* \}$.
 \end{definition}
\section{Generalized Nash Equilibrium seeking}
\label{sec:GNE_seeking}
We now turn our attention to the derivation of a distributed algorithm to find a GNE of the problem in \eqref{eq:generalized_game}.  Let us formulate the following feasibility assumption: 
\begin{assumption} \label{as:slater}
	The set $\bOmega$ is non-empty and it satisfies Slater's {constraint qualification \cite[Eq. 27.50]{bauschke_convex_2017}. }
\end{assumption}
{$\bOmega$ in Definition \ref{def:gne} is compact and convex because defined by linear equations.} Furthermore, the local cost functions are convex, as formalized next:
\begin{lemma} \label{le:cost_is_convex}
 Let Assumption \ref{as:ell}, \ref{as:local_cost} hold. For each $i\in\mc I$,	$J_i\left(\omega_i, \bomega_{-i}\right)$ is convex in $\omega_i\in\Omega_i$ for all $\bomega_{-i}\in \bigtimes_{j\in\mc I_{-i}} \Omega_j$.
\end{lemma}
Under  Assumption \ref{as:slater} and Lemma \ref{le:cost_is_convex}, we conclude that a GNE exists \cite[Prop. 12.11]{palomar_convex_2009} if the game  mapping
\begin{equation}
	\label{eq:pseudogr}
	F(\bomega) :=  \mathrm{col}\left(  \nabla_{\omega_i}J_i(\omega_i, \bomega_{-i})\right) _{i\in\mc I} 
\end{equation} 
is monotone. Monotonicity of the game mapping is one of the mildest conditions under which effective GNE seeking algorithms can be derived. The authors of \cite{bakhshayesh_decentralized_2021} show that monotonicity of the game defined using $\ell_{(a,b)}^{\text{BPT}}$ in \eqref{eq:bpt_latency} for all $(a,b)$ holds if enough non-controlled vehicles populate the roads \cite[Eq. 18]{bakhshayesh_decentralized_2021}, with the caveat that this number must increase proportionally with the number of controlled vehicles. This {might not be reasonable if} the share of controlled vehicles is large. In the following, we derive a milder condition. 

\subsection{Monotonicity of the game}
To study the monotonicity of the proposed formulation, let us define for each road $(a,b)$ and instant $t$ the operators
\begin{align}
	\label{eq:def_F_ab}
	\begin{split}
		&F_{(a,b),t}(\bs{M}_{t, (a,b)} )\hspace{-2pt} :=\hspace{-2pt} \mathrm{col}( J_{(a,b)}'(\cdot, \bs M^{-i}_{t,(a,b)} )|_{M^i_{t, (a,b)}})_{i\in\mc I}
	\end{split}
\end{align} 
where we compute 
	\begin{align}
		\label{eq:derivative_element_J}
		\begin{split}
			J_{(a,b)}'(\cdot, \bs M^{-i}_{(b,a),t})|_{M^i_{t, (a,b)}} &= \ell_{(a,b)}( \sigmaM_{(a,b), t})  +  \\
			& \tfrac{1}{N} M^i_{t, (a,b)} \ell'_{(a,b)}( \sigmaM_{(a,b), t} ).\\
		\end{split}
	\end{align}
We now link the monotonicity of $F$ to that of each $F_{(a,b),t}$.
\begin{lemma} \label{le:monotonicity_decomposition}
The operator $F$ in \eqref{eq:pseudogr} is monotone if $F_{(a,b),t}$ in \eqref{eq:def_F_ab} is monotone for each $(a,b)$ and $t$.
\end{lemma}
For a particular class of $\ell_{(a,b)}$ {(which includes $\ell^{\text{BPT}}_{(a,b)}$ in \eqref{eq:bpt_latency}), we find the following monotonicity condition:}
\begin{lemma} \label{le:monotonicity_condition}
	Let $\ell : \R_{\geq 0} \to \R_{\geq 0}$ {be defined as}
	 \begin{equation} \label{eq:l_monomial}
		\ell(\sigma) = \tau +  \tfrac{k}{\xi+1}(\sigma+\zeta)^{\xi+1}  
	\end{equation}
for some $\tau, k, \xi, \zeta \in\R_{\geq 0}$. Let
	\begin{equation}
		\label{eq:operator_T}
		 T(\by) := \mathrm{col} \left( \ell( \mathrm{avg}(\by)) + \tfrac{1}{N} y_i \nabla \ell( \mathrm{avg}(\by)  \right)_{i\in\mc I}.
	\end{equation}
	Then, $T$ is monotone on $[0,1]^N$ if 
	\begin{equation}
		\label{eq:monotonicity_condition}
		\zeta\geq \max\left( \tfrac{\xi^2-8}{8N}, \tfrac{\xi-2}{2N} \right).
	\end{equation} 
\end{lemma}
\begin{remark}
\eqref{eq:monotonicity_condition} is satisfied for any $\zeta$ whenever $\xi \leq 2$.  
\end{remark}
\begin{remark} {Let us compare the condition in \eqref{eq:monotonicity_condition} with the {previously known monotonicity condition derived in} \cite{bakhshayesh_decentralized_2021} for the latency function $\ell_{(a,b)}^{\text{BPT}}$ with $\xi=3$. } Following \eqref{eq:monotonicity_condition} the game is monotone if  $\zeta_{(a,b)}\geq\frac{1}{2N}$ for all $(a,b)$, which is satisfied if at least $\frac{V}{2}$ uncontrolled vehicles traverse each road. By contrast, taking into account that \cite{bakhshayesh_decentralized_2021} considered $V=1$, the bound in \cite[Eq. 18]{bakhshayesh_decentralized_2021} requires $\frac{3NV}{8}$ uncontrolled vehicles on each road when translated to our setting. {By applying Lemma \ref{le:monotonicity_condition}, one can then find more general conditions for the convergence of  \cite[Alg. 1]{bakhshayesh_decentralized_2021} than the ones specified in \cite[Thm. 1]{bakhshayesh_decentralized_2021}. }
\end{remark}
In view of  Lemma \ref{le:monotonicity_condition}, let us assume the following:
\begin{assumption} \label{as:monotonicity_condition} {For all} $(a,b)\hspace{-2pt}\in \hspace{-1pt}\mc E$, $\ell_{(a,b)}$ in \eqref{eq:def_scalar_cost} is in the form
	$$\ell_{(a,b)}(\sigma) =  \tau_{(a,b)} +  \tfrac{k_{(a,b)}}{\xi+1}(\sigma+\zeta_{(a,b)})^{\xi+1}$$
	where $\xi$, $\tau_{(a,b)}$, $k_{(a,b)}\in\R_{\geq 0}$ and  $\zeta_{(a,b)}$ satisfies \eqref{eq:monotonicity_condition}.
\end{assumption}
 Assumption \ref{as:ell} is implied by Assumption \ref{as:monotonicity_condition}. For each $(a,b)\in \mc E$, $t\in \mc T$, $F_{(a,b),t}$ is in the form in \eqref{eq:operator_T}, as can be seen by substituting  \eqref{eq:derivative_element_J} in \eqref{eq:def_F_ab}. Thus, $F_{(a,b),t}$ is monotone on $[0,1]^N$ by  Lemma \ref{le:monotonicity_condition}.  As $\bs\Omega\subset[0,1]^{Nn_{\omega}}$ by Lemma \ref{le:M_reformulation}, the following result is immediate by Lemma \ref{le:monotonicity_decomposition}:
\begin{lemma} \label{le:game_is_monotone}
	Under Assm. \ref{as:monotonicity_condition}, $F$  in \eqref{eq:pseudogr} is monotone on $\bOmega$.
\end{lemma}
{Lemma \ref{le:game_is_monotone} is fundamental for guaranteeing the convergence of the GNE-seeking algorithm proposed in Section \ref{sec:GNE_seeking:algorithm}. }
\subsection{Semi-decentralized equilibrium seeking}\label{sec:GNE_seeking:algorithm}
 To solve the game in \eqref{eq:generalized_game}, we focus on the computation of a variational GNE (v-GNE) \cite[Def. 3.10]{facchinei_generalized_2010}, that is, the subset of GNEs which satisfy the KKT conditions
 \begin{equation}
 	\label{eq:kkt}
 	\begin{bmatrix}
 		\bomega \\ \lambda
 	\end{bmatrix}\hspace{-2pt}\in \hspace{-1pt} \zer\left(  \begin{bmatrix}
 		\bigtimes_{i\in\mc I} \nc_{\Omega_i}(\omega_i) \hspace{-1pt}+\hspace{-1pt} F(\bomega)\hspace{-1pt}+\hspace{-1pt} \col(A_i^\top  \lambda)_{i} \\
 		\nc_{\bb R^{|\mc E|}_{\geq 0}}( \lambda) \hspace{-1pt}-\hspace{-1pt} A  \bomega + b
 	\end{bmatrix}\right)
 \end{equation}
 where $\lambda \in \R^{|\mc E|}_{\geq 0}$ is the dual variable associated to the shared constraints in \eqref{eq:shared_constr}. The v-GNEs have desirable characteristics of fairness between agents and there exist several efficient algorithms for their computation. In particular, we adopt the Inertial Forward-Reflected-Backward (I-FoRB) algorithm \cite{belgioioso_semi-decentralized_2023}, for its convergence speed and low computational complexity. The I-FoRB algorithm converges in the general class of (non-strictly) monotone games. On the contrary, the algorithm proposed in \cite{bakhshayesh_decentralized_2021} converges only if the game is strongly monotone, thus an additive quadratic cost is necessary, which is not needed in our model. The agents perform a \emph{reflected} projected-gradient descent of the Lagrangian function with an inertial term \eqref{eq:primal_update}. Then, the agents communicate the primal variable and auxiliary variables $\bs{d}$ to the aggregator. In turn, the aggregator updates the aggregate variable and the dual variable via a reflected dual ascent with inertia  \eqref{eq:dual_update_1} and communicates them to the agents. {We now state the main result of this section, where we denote by $L_i^f$ the Lipschitz constant of $f_i$ for all $i\in\mc I$ (which exist following Assumption \ref{as:local_cost} and the compactness of $\bOmega$): 
 \begin{proposition} \label{prop:main}
  Let Assumptions \ref{as:local_cost}, \ref{as:slater}, \ref{as:monotonicity_condition} hold and let 
  \begin{align*}
  &	L_{(a,b)} \geq {\tfrac{k_{(a,b)}}{N} ((1 + \zeta_{(a,b)})^{\xi} + \xi(1 + \zeta_{(a,b)})^{\xi-1})}\\
  &L\geq {\textstyle\max_i(L^f_i) +} \textstyle\max_{(a,b)\in\mc E} (L_{(a,b)}) \\
  &  \theta \in [0,\tfrac{1}{3}) , ~~~ \delta >2L/(1-3\theta).
  \end{align*}
  Then, $(\bomega^{(k)}, \lambda^{(k)})_{k\in\N}$ generated by Algorithm \ref{alg:FoRB} with stepsizes
  \begin{align*}
  	\begin{split}
  &	0<\alpha_i \leq (\|A_i\| + \delta)^{-1} ~~~~ \text{for all}~i\in\mc I \\
  &	0<\beta \leq N(\textstyle\sum_{i=1}^N\|A_i\| + \delta)^{-1}
  \end{split}
  \end{align*}
converges to $\col(\bomega^*, \lambda^*)$ where $\bomega^*$ is  a v-GNE of \eqref{eq:generalized_game}. 
 \end{proposition}
 	\begin{algorithm}[t!]
 	\caption{I-FoRB GNE seeking  for traffic routing}
 	\label{alg:FoRB}
 	\textbf{Initialization.} $\forall i\in\mc I$ set $\omega_i^{(0)}, \omega_i^{(1)} \in \R^{n_{\omega}}$; $\lambda^{(1)},\lambda^{(0)} \in \bb R_{\geq 0}^{|\mc E|}$.\\
 	\textbf{For $k\in\N$}:
 		\begin{enumerate}[leftmargin=*] 
 			\item Each agent $i \in \mc I$  receives $\sigma^{(k)}$, $\lambda^{(k)}$ and computes:
 			{
 				\begin{subequations}
 					\begin{align}
					 r_i^{(k)} =& 2\nabla_{\omega_i} J_i(\omega_i^{(k)}, \bomega_{-i}^{(k)}) - \nabla_{\omega_i} J_i(\omega_i^{(k-1)}, \bomega_{-i}^{(k-1)})\\
 					\omega_i^{(k+1)} =& \proj_{\Omega_i}  ({\omega}_i^{(k)} - \alpha_i(r_i^{(k)} + A_i^{\top} \lambda^{(k)}) \nonumber \\
 					 &+ \theta(\omega_i^{(k)} - \omega_i^{(k-1)})) \label{eq:primal_update} \\ 
 					d_i^{(k+1)} =& 2 A_i \omega_i^{(k+1)} - A_i\omega_i^{(k)} - b_i. \label{eq:dual_update}
 				\end{align}
 			\end{subequations}}
 			\item The aggregator receives $(\bs{\omega}^{(k+1)}, \bs{d}^{(k+1)})$ and computes:
 		\end{enumerate}
 		\begin{subequations}
			\begin{align}
	 				\sigma^{(k+1)} &= \mathrm{avg}(\bs\omega^{(k+1)})\\
					\lambda^{(k+1)} &= \proj_{\mathbb{R}_{\geq 0}^{|\mc E|}} (\lambda^{(k)} + \beta \mathrm{avg}(\bs d^{(k+1)}) +\theta(\lambda^{(k)} - \lambda^{(k-1)}) ) \label{eq:dual_update_1}
			\end{align}
 		\end{subequations}
 \end{algorithm}
\section{Receding horizon game formulation}
\label{sec:receding_horizon_GNE}
Due to the constraint in \eqref{eq:final_dest_constr}, the problem in \eqref{eq:generalized_game} admits a solution only if, for all $i$, the destination $d_i$ is reachable from the starting node $b_i$ in $T+1$ steps.  In common applications, choosing a  large enough $T$ that guarantees feasibility might lead to a heavy computational burden. In this section, we propose an alternative formulation of the problem in \eqref{eq:generalized_game} without the constraint in \eqref{eq:final_dest_constr} as $N$ coupled Finite Horizon Optimal Control Problems (FHOCPs), which we label Finite Horizon Multi-Stage Game (FHMSG). Inspired by the Model Predictive Control (MPC) literature, we propose to repeatedly solve the FHMSG in receding horizon and to iteratively apply the first input of the computed sequence. The agents are directed to their destinations by means of a terminal cost which penalizes the distance from their destination. Remarkably, the introduction of a terminal cost is reminiscent of the classic stability requirements of MPC \cite[Sec. 2]{rawlings_model_2017}.
\subsection{Equivalent finite horizon optimal control problem}
Let us formalize the game under consideration, parametrized in the initial distribution $\rho_{\text{in}}^i\in\Delta^{|\mc N|}$:
	\begin{align}
		\label{eq:game2}
			\text{for all} ~ i\in\mc I: ~\underset{\omega_i \in \mc {Y}_{i}}{\min}  & J_i(\omega_i, \bomega_{-i})
	\end{align}
where $\mc Y_i:=   \left\{\omega\in {\R^{n_{\omega}}} | \eqref{eq:M_constr_1}, \eqref{eq:M_constr_2}, \eqref{eq:M_constr_3}, \rho_1^i =\rho_{\text{in}}^i \right\}$. We  emphasize that we do not include the constraint in \eqref{eq:shared_constr}: due to the probabilistic control action, an unlucky realization might lead the constraint \eqref{eq:shared_constr} to be unfeasible at the successive time steps.  Instead, $\mc Y_i$ is non-empty for any $\rho_{\text{in}}^i$. {The exclusion of \eqref{eq:shared_constr} renders the problem in \eqref{eq:game2} a (non-generalized) game.}
In this section, we show the equivalence of the problem in \eqref{eq:game2} to a FHOCP. As a first step, we rewrite the equations defining $\mc Y_i$ as the state-space representation of a constrained linear system. 
We define the desired distribution $\rhoeq^i$ as {the one where every vehicle is at their destination, that is,} $ [\rhoeq^i]_a = \delta_{d_i}(a) $, where $\delta_{d_i}$ is a Kronecker delta centered in $d_i$, and {the associated equilibrium input} $\ueq^i: =\col(\delta_{d_i}(a)\delta_{d_i}(b))_{(a,b)\in\mc E}$, that is, the vector of edge transitions associated to {remaining in the destination node $d_i$} with probability $1$.
We define the states $x_t^i \in\R^{\nx}$ and inputs $u_t^i\in\R^{\nu}$ {such that the origin coincides with the desired distribution and equilibrium input, that is:}
\begin{align}
	&x_t^i := \rho_t^i - \rhoeq^i \label{eq:state_def} \\
	&	u_t^i := \col(M^i_{t, (a,b)} )_{\substack{(a,b) \in \mc E }} - \ueq^i. \label{eq:input_def}
\end{align}
We define the selection vectors $\Se^{(a,b)}\in\R^{\nu}$ for all $(a,b)\in\mc E$ such that $(\Se^{(a,b)})^{\top} (u_t^i+\ueq^i) = M^i_{t, (a,b)}$, as well as 
\begin{align*}
	&B:=\col( \textstyle\sum_{a: (a,b)\in\mc E} 	(\Se^{(a,b)})^{\top} )_{b\in\mc N} \\
	&P:=\col( \textstyle\sum_{b: (a,b)\in\mc E} 	(\Se^{(a,b)})^{\top} )_{a\in\mc N}.
\end{align*}
It can be verified that  $ B\ueq^i =  P\ueq^i = \rhoeq^i  $ and thus, by substituting the definitions of $B$ and $P$:
\begin{subequations}
	\label{eq:input_state_constraints}
	\begin{align}
\label{eq:constrained_dynamics:dynamics}	\eqref{eq:M_constr_1} &\Leftrightarrow  x_{t+1}^i
	= B u_t^i \\
\label{eq:input_state_constraint_2}	\eqref{eq:M_constr_2} &\Leftrightarrow x_{t}^i
	 =P u_t^i.
	\end{align}
\end{subequations}
By substituting \eqref{eq:input_def} in \eqref{eq:constrained_dynamics:dynamics}, the desired state $x_t^i=0$ is an equilibrium {if the action $M_{t, (d_i,d_i)}^i=1$ is applied.} We then make sure that {such action is cost-free for all agents (and thus, trivially, a NE strategy) when the initial state is the origin by assuming that the self-loops have no traversing cost.} {Another restriction on the cost functions is supported by the following lemma, which states that $\ell_{(a,b)}$ must be affine for the game in \eqref{eq:generalized_game} to be potential \cite[Sec. 2]{monderer_potential_1996}. This property is crucial to rewrite \eqref{eq:game2} as a single optimization problem.
 	\begin{lemma}
 		\label{le:congestion_must_affine}
 		Let Assm. \ref{as:ell} hold. The game in \eqref{eq:game2} is potential if and only if $\ell_{(a,b)}$ in \eqref{eq:def_scalar_cost} is affine for each $(a,b)\in\mc E$.
 	\end{lemma}} In view of these requirements, we formulate the following assumption (which implies Assumption \ref{as:monotonicity_condition}):
\begin{assumption}
 	\label{as:linear_cost}
	For each $(a,b)\in\mc E$, the congestion function $\ell_{(a,b)}$ is of the form
	$\ell_{(a,b)}(\sigma)= \tau_{(a,b)} + k_{(a,b)} \sigma$, with
		$\tau_{(a,a)} = 0$, $k_{(a,a)} = 0$ and
			$\tau_{(a,b)} > 0$ for all $a\in \mc N$,  $b\in\mc N \setminus \{a\}$.	
\end{assumption}
 Linear(ized) edge traversing costs in traffic routing problems are considered in \cite{angelelli_system_2021, calderone_markov_2017} and in the numerical analysis of \cite{bakhshayesh_decentralized_2021}.  Naturally, Assumption 
 \ref{as:linear_cost} implies that an optimal policy for each agent is to remain at their initial state. We then impose that the agents cannot  choose to remain still in a node, unless that node is their destination:
\begin{equation}
	\label{eq:self_loop_constr}
	(\Se^{(a,a)})^{\top} u_{t}^i = 0 ~~~ \forall~a\in\mc N \setminus\{d_i\}, ~i\in\mc I,~t\in\mc T.
\end{equation}
 {We can compactly  rewrite $\mc Y_i$ with the additional constraint \eqref{eq:self_loop_constr} as the dynamics in \eqref{eq:constrained_dynamics:dynamics} with initial state $x_1^i = \rho_{\text{in}}^i - \rho_{\text{eq}}^i  $ and constraint set $\mathbb{Z}_i$ defined as}
{ \begin{align*}
		&\mathbb{X}_i :=  \Delta^{|\mc N|} - \{\rhoeq^i\}, \\
		&\mathbb{U}_i := \{u\in \R_{\geq 0}^{|\mc E|} - \{\ueq^i\}|\Se^{(a,a)\top}u =0 ~\forall a\in\mc N \setminus \{d_i\} \}, \\
		& \mathbb{Z}_i:= \{ (x,u) \in \mathbb{X}_i\times \mathbb{U}_i | u\in \mathrm{null}(P) + \{P^{\dagger}x\}  \}.
	\end{align*}
As a next step towards the formulation of the multi-stage decision problem, we rewrite the cost $J_i$ in terms of stage cost, that is, as a sum of terms which only depends on the variables at time step $t\in\mc T$. We make the following technical assumption, which is typical of multi-stage decision problems:
\begin{assumption}
	\label{as:local_cost_separable}
	The local cost $f_i(\omega_i)$ is separable in $t$, that is, $f_i(\omega_i) = \sum_{t\in\mc T} f^S_{i}(x_t^i, u_t^i)  + f_{i}^F (x_{T+1}^i)$. Furthermore, $f^S_{i}$ and $f_{i}^F$ are non-negative and $f^S_{i}(0, 0)=f_{i}^F (0)=0$.
\end{assumption}
Let us collect the network parameters in $\bar{\tau}: = \sum_{(a,b)} \tau_{(a,b)}\Se^{(a,b)}$ and
$ C:= \textstyle\sum_{(a,b) } k_{(a,b)} \Se^{(a,b)}(\Se^{(a,b)})^{\top}$. It can then be shown (see Appendix \ref{appx:derivation_rewritten_cost}) that under Assumptions \ref{as:linear_cost}, \ref{as:local_cost_separable}, $J_i$ in \eqref{eq:game2} can be written {in terms of stage costs} as:
{\begin{align}
	\begin{split}
	\label{eq:cost_rewritten}
	& J_i(x^i_1,(\bu_{t})_{t\in\mc T}  ) =  {f}_{i}^F (x_{T+1}^i)+  \\
	&\qquad\textstyle\sum_{t\in\mc T} {f}^S_{i}(x_{t}^i, u_t^i ) + \bar{\tau}^{\top} u_t^i  +\mathrm{avg}(\bu_{t})^{\top}C u_t^i.
	\end{split}
\end{align}}
  We then formulate the FHMSG $\mc G(\bx_{1})$ for any $x^i_1 \in \mathbb{X}_i$: 
\begin{subequations}
	\label{eq:MPC_game}
	\begin{empheq}[left={{\footnotesize\forall} i: \empheqlbrace}]{align}
		\min_{\substack {(u_{t}^i)_{t\in\mc T} } } &  J_i(x^i_1, (\bu_{t})_{t\in\mc T}  )  \label{eq:MPC_game:cost}\\ 
		\text{s.t.}~~~ &  (x^i_t, {u}_{t}^i) \in \mathbb{Z}_i ~~~~ \forall t\in\mc T,
	\end{empheq}
\end{subequations}
which is equivalent to \eqref{eq:game2} with the additional constraint in \eqref{eq:self_loop_constr}.} Next, we show that the game in \eqref{eq:MPC_game} is equivalent to a FHOCP. {Recall that $\bx_t=\col(x_t^i)_{i\in\mc I}$ and note that $\bx_t$ evolves according to the collective dynamics $\bx_{k+1} = (I_N\otimes B) \bu_k$.} Define the constraint sets for the collective states and inputs:
\begin{align*}
	&\mathbb{X} := \{\bx | {x_i} \in \mathbb{X}_i ~~ \text{for all}~i\} \\
	&\mathbb{U} := \{\bu | {u_i} \in \mathbb{U}_i ~~ \text{for all}~i\} \\
	&\mathbb{Z} := \{(\bx,\bu) | {(x_i, u_i)} \in \mathbb{Z}_i ~~ \text{for all}~i\}.
\end{align*}
{In the following Lemma \ref{le:potential_of_MPC_game} we find a common control objective $p$ that the agents unknowingly, but willingly, aim at minimizing when solving the game $\mc G$: }
\begin{lemma}
	\label{le:potential_of_MPC_game}
	{For any $\bs{x}_1\in\mathbb{X} $, the pseudogradient of $\mc{G}(\bs{x}_1)$ in \eqref{eq:MPC_game} is equal to $\nabla_{\bu} {\potential}$, where}
	\begin{subequations}
	\label{eq:potential_of_game}
	\begin{align}
		&\potential^S(\bx_t,  \hspace{-1pt}\bu_t) \hspace{-2pt} :=    \hspace{-2pt} \tfrac{1}{2N}\|\bu_t\|^2_{(I_N   +  \bs {1}\bs {1}^{\top} ) \otimes C}   \hspace{-2pt}+ \hspace{-2pt} \textstyle\sum_{i\in\mc I}   \hspace{-1pt}{f}^S_{i}(x^i_t, u^i_t)  \hspace{-2pt}+  \hspace{-2pt} \bar{\tau}^{\top} \hspace{-1pt}u^i_t   \label{eq:potential_of_game:stage}\\
		&	 \potential^F(\bx_{T+1}) := \textstyle\sum_{i\in\mc I} {f}_i^F(x_i) \label{eq:potential_of_game:final}\\
		&{\potential}(\bx_{1}, (\bu_{\tau})_{\tau\in\mc T})  \hspace{-2pt} :=  \hspace{-2pt} \potential^F(\bx_{T+1})+ \textstyle\sum_{t\in\mc T} \potential^S(\bx_t, \bu_t). \nonumber
	\end{align}
\end{subequations}
\end{lemma}
{The potential function $p$ allows us to conclude that,} by Lemma \ref{le:potential_of_MPC_game} and  \cite[Theorem 2]{scutari_potential_2006}, a NE  of $\mc G(\bx_{1})$  is a solution to the FHOCP $\mc O(\bx_{1})$, defined as: 
\begin{subequations}
	\label{eq:MPC_opt_problem}
	\begin{empheq}[left={\empheqlbrace}]{align}
		\min_{({\bu}_t)_{t\in\mc T}  } ~~~&{\potential}(\bx_{1}, (\bu_\tau)_{\tau\in\mc T} )  \label{eq:MPC_opt_problem:cost}\\
		\text{s.t.} ~~~&  (\bx_t, \bu_t) \in \mathbb{Z}~~~\forall t\in\mc T.
	\end{empheq}
\end{subequations}
{We  now show the  asymptotic stability of the receding horizon solution of \eqref{eq:MPC_opt_problem}, and in turn of \eqref{eq:MPC_game}, via standard MPC results. 
\subsection{Stability of receding horizon Nash equilibrium control}
At every time-step, the agents apply the first input corresponding to a NE of the game in \eqref{eq:MPC_game}. This is formalized via the following control actions:
\begin{equation}
	\label{eq:control_action}
	\kappa_i \hspace{-1pt}:  \hspace{-1pt}y \hspace{-1pt}\mapsto \hspace{-1pt} u_1^{i*} ~\text{where}~\col(u_t^{i*})_{t\in\mc T,i\in\mc I}~\text{is a NE of}~\mc G( y).
\end{equation}
 Intuitively, $\kappa_i$ leads the $i$-th agent to the desired equilibrium if the agents have an high enough incentive to approach their destinations. For this purpose, let us assume that each agent knows a path to its destination, formalized by the mappings $(\NP{i})_{i\in\mc I}: \mc N \to \mc N $ with the following characteristics:
\begin{align*}
		&\NP{i}(d_i) = d_i; ~(a,\NP{i}(a)) \in\mc E;\\
		& \exists~T_i^P\in \mathbb{N}~ \text{such that} ~\NP{i}^t(a):= \underbrace{\NP{i} \circ ... \circ \NP{i}}_{t ~\text{times}}(a) = d_i \\
		& \text{for all}~ a\in\mc N, ~ t\geq T_i^P.
\end{align*}
An example is the shortest path computed with edge weights $\bar{\tau}$. {We define the traversing time of the next edge along the known path and for the whole path starting from each node:}
	\begin{alignat*}{3}
	\tausp{i}\in \R^{|\mc N|}_{\geq 0}; ~~& [\tausp{i}]_a := \tau_{(a, \NP{i}(a)) }, \\ 
	\musp{i}\in \R^{|\mc N|}_{\geq 0}; ~~& [\musp{i}]_a := \textstyle\sum_{t=0}^{\infty} [\tausp{i}]_{\NP{i}^t(a)},	\end{alignat*}
and the following {auxiliary} input, designed such that every vehicle takes the next edge of the known path:
\begin{align}
	\label{eq:le:lyap_descent:input_choice}
	\begin{split}
		& \usp{i}:\mathbb{X}_i\to \mathbb{U}_i ~ \text{for all}~ i~\text{such that} \\
		&	\Se^{(a,b)^\top} \usp{i}(x_i)= \begin{cases}[x_i]_a - \delta_{d_i}(a) &\text{if}~ b=\NP{i}(a) \\
			0 &\text{if}~ b\neq \NP{i}(a). \end{cases}
	\end{split}
\end{align}
We then postulate the following technical assumption, which encodes the fact that each agent evaluates {the distance of the final state} from the destination by means of the known path:
\begin{assumption}
	\label{as:cost_eval_wrt_shortest_path} The local costs satisfy Assumption \ref{as:local_cost_separable} with 
	\begin{align*}
		& {f}^F_{i}(x) = \sigma_i^F((\musp{i})^{\top} x ) \\ 
		& {f}_{i}^S(x,\usp{i}(x)) \leq \sigma_i^S((\tausp{i})^{\top} x ),
	\end{align*}
where $ \sigma_i^F$  is a $m_F$-strongly monotone and $\sigma_i^S$ is a $L_{\text{S}}$-Lipschitz continuous functions  for all $i$, with $ \sigma_i^F(0)=\sigma^S_i(0)=0$.
\end{assumption}
For example, Assumption \ref{as:cost_eval_wrt_shortest_path} is satisfied by $f_i^F(x) \hspace{-1pt}= \hspace{-1pt} \gamma_1 (\musp{i})^\top x$, with $\gamma_1\hspace{-1pt}>\hspace{-1pt}0$ and  $ f_i^S(x, u)  \hspace{-1pt}=\hspace{-1pt}  \gamma_2 (\tausp{i})^{\top} \hspace{-1pt}x$, with $\gamma_2\geq 0$. 
{In the main result of this section we show that, if the agents have a high enough incentive to reach the destination (encoded by the strong monotonicity constant of the terminal cost $m_F$), then the system in \eqref{eq:constrained_dynamics:dynamics} controlled by the receding-horizon NE control $(\kappa_i)_{i\in\mc I}$ defined in \eqref{eq:control_action} asymptotically reaches the origin} (that is, the state at which every vehicle is at its destination with probability 1).} {The proof follows from the equivalence between \eqref{eq:MPC_game} and \eqref{eq:MPC_opt_problem}. Specifically, we show that $p^F$ is a control Lyapunov function under control action $u^{\mathrm{kp}}$ for the collective system whose map from input to state is $I_N \otimes B$ (cf. \eqref{eq:constrained_dynamics:dynamics}). We then apply a known result in MPC theory \cite[Theorem 2.19]{rawlings_model_2017} to conclude the asymptotic stability.}
\begin{theorem}
	\label{thm:main_receding_hor}
	Denote $\bar{k}:= \max_{(a,b)} (k_{(a,b)})$ and $\tau_{\text{min}}:= \min_{(a,b), a\neq b} \tau_{(a,b)}$. Under Assumptions \ref{as:strongly_connected},\ref{as:local_cost},\ref{as:linear_cost}--\ref{as:cost_eval_wrt_shortest_path} and if
	\begin{equation}
		\label{eq:condition_gamma_terminal}
		m_F\geq 1+ L_{\text{S}} + \tfrac{\bar{k}(N+1)}{2N\tau_{\mathrm{min}}},
	\end{equation} 
	then  the origin is asymptotically stable for the systems $x_{t+1}^i = B\kappa_i(\bx_{t})$ for all $i\in\mc I$, with $\kappa_i$ as in \eqref{eq:control_action}.
\end{theorem}
Let us present the resulting approach in Algorithm \ref{alg:rec_hor}.

 	\begin{algorithm}[t!]
	\caption{Receding horizon NE seeking for traffic routing}
	\label{alg:rec_hor}
	\textbf{Initialization.} Set $\rho_1^i$ as in \eqref{eq:M_constr_4} for each $i\in\mc I$.\\
	\textbf{For } $\tau \in\mathbb{N}$: 
	\begin{enumerate}[leftmargin=*] 
			\item		 
			\textbf{Agents control computation}: 
		\begin{enumerate}
			\item A NE  of $\mc G(\bs\rho_{\tau})$
			$$([\mathrm{row}({M}^{i*}_{t, (a,b)})_{\substack{(a,b)\in\mc E ,  t\in\mc T}}, \mathrm{row}(\rho^{i*\top}_{t})_{t\in\mc T^+} ]^{\top})_{i\in\mc I}$$ is computed using Algorithm \ref{alg:FoRB}, where each $(\Omega_{i})_{i\in\mc I}$  in \eqref{eq:primal_update} is substituted with $\{\omega_i\in\mc Y_i| \eqref{eq:self_loop_constr} ~\text{holds}\}$, $\lambda^{(1)}=\bs 0$ and the dual update \eqref{eq:dual_update_1} is ignored.
			\item Each agent $i$ computes $\Pi_1^{i*}$ as in \eqref{eq:inverse_transform}. 
		\end{enumerate}
		\item \textbf{Vehicles node update}:\\
				 For all $v\in\{1,...,V\}$, $i\in\mc I$ draw $s^{i,v}_{\tau+1}\in\mc N$ from the probability distribution $\col([\Pi_1^{i*}]_{(b, s^{i,v}_{\tau})})_{b\in\mc N}$
			\item \textbf{Agents state update}:\\
				 Each agent updates the empirical distribution:
				{
				\begin{align*}
					&\pi_{n,i} = | \{ v\in\{1,...,V\}~ \text{s.t.}~ s^{i,v}_{\tau+1} = n \} |~ \text{for all}~n\in\mc N \\
					&\rho_{\tau+1}^i=\col(  \pi_{n,i}  /V)_{n\in\mc N}
				\end{align*}}
		\end{enumerate}
\end{algorithm}

\section{Numerical study$^1$}
\label{sec:numerical}
\let\thefootnote\relax\footnote{$^1$Code available at \protect\url{https://github.com/bemilio/MDP_traffic_nonlinear}}
We {study the behavior of} Algorithm \ref{alg:FoRB} {on multiple randomly generated simple scenarios, in order to better observe the characteristics of the solution.} We implement a {randomly generated} directed graph with $12$ nodes and $27$ edges for $N=8$ agents. We consider the case where the agents only take the traversing time into account when choosing the road, and thus we} set $f_i\equiv 0$ for all $i$. {We set every road to have the same length and capacity, by considering}  $\ell_{(a,b)}$ to the BPT latency function in \eqref{eq:bpt_latency} with $\tau_{(a,b)}=0.1$, $\zeta_{(a,b)}=\frac{1}{N}$, $c_{(a,b)} = 0.1$ for all $(a,b)\in\mc E$ and $\xi=3$. The road limit in \eqref{eq:shared_constr} is defined as $\overline{c}_{(a,b)} = 0.2$. We solve the problem in \eqref{eq:generalized_game} for $100$ random initial states and destinations of the agents. The solution to the game in \eqref{eq:generalized_game} is then compared to  the routing obtained by the shortest path with no traffic information, that is, with edge weight $\tau_{(a,b)}$ for all $(a,b)\in\mc E$. We can conclude from Figure \ref{fig:2_comparison_congestions} that the baseline solution tends to overcrowd some roads (cf. edge 6) and underutilize others (cf. edge 3), while the proposed GNE routing, which exploits traffic information, obtains a more uniform usage of the network. {In   Figure \ref{fig:4_comparison_expected_congestion}, we show the normalized approximation error for the travel time computed using \eqref{eq:first_order_approx}. The approximation error for each link $(a,b)\in\mc E$ and instant $t$ is computed as $ |\ell_{(a,b)}(\sigma_{(a,b),t}^{\mathrm{M}}) - \ell_{(a,b)}(\hat{\sigma}_{(a,b),t}^{\mathrm{M}})|$, where $\sigma^{\mathrm{M}}_{(a,b),t}$ is the expected road occupation defined in \S \ref{sec:objective_and_constraints} and $\hat{\sigma}^{\mathrm{M}}_{(a,b),t}$ is the realized number of vehicles on $(a,b)$ at time $t$ divided by $NV$.}  Evidently, increasing the population size reduces the approximation error. We then apply Algorithm \ref{alg:rec_hor} for $\tau\in\{1,...,10\}$ with the terminal cost $f_i^F(x) = \gamma (\musp{i})^{\top} x$ and $\gamma$ as in the right hand side of \eqref{eq:condition_gamma_terminal}, {which ensures that the assumptions of Theorem \ref{thm:main_receding_hor} are satisfied.} The results are compared to the pre-computed open-loop solution of problem \eqref{eq:generalized_game} without the constraint in \eqref{eq:shared_constr}, {denoted by the $\infty$ horizon, in terms of the relative total traversing time reduction with respect to the shortest-path solution without traffic information.} Figure \ref{fig:2_multiperiod_advantage}  shows that the traversing time experienced is reduced with respect to the shortest path solution, and this advantage increases with the time horizon. In a practical sense, the results in Figures \ref{fig:2_comparison_congestions}, \ref{fig:2_multiperiod_advantage} show that the availability of real-time traffic information allows to reduce the congestion on heavily utilized links and the total traversing time experienced by the drivers. 

\section{Conclusion}   
Traffic routing of multiple vehicles can be modelled as an aggregative game with mixed strategies using a first-order approximation of the latency function. The approximation error decreases as the number of controlled vehicles increases. The particular structure of the road latency function guarantees the monotonicity of the game under mild conditions, allowing for solution via existing equilibrium-seeking algorithms. If the latency function is linear, then the game can be solved in receding horizon whenever the local objective functions satisfy a set of conditions inherited from the MPC literature. {Numerical simulations show that the proposed solution reduces the overall network congestion and traversing time, compared to the optimal routing computed without traffic information.} 

\begin{figure}
	\includegraphics[width=0.93\columnwidth]{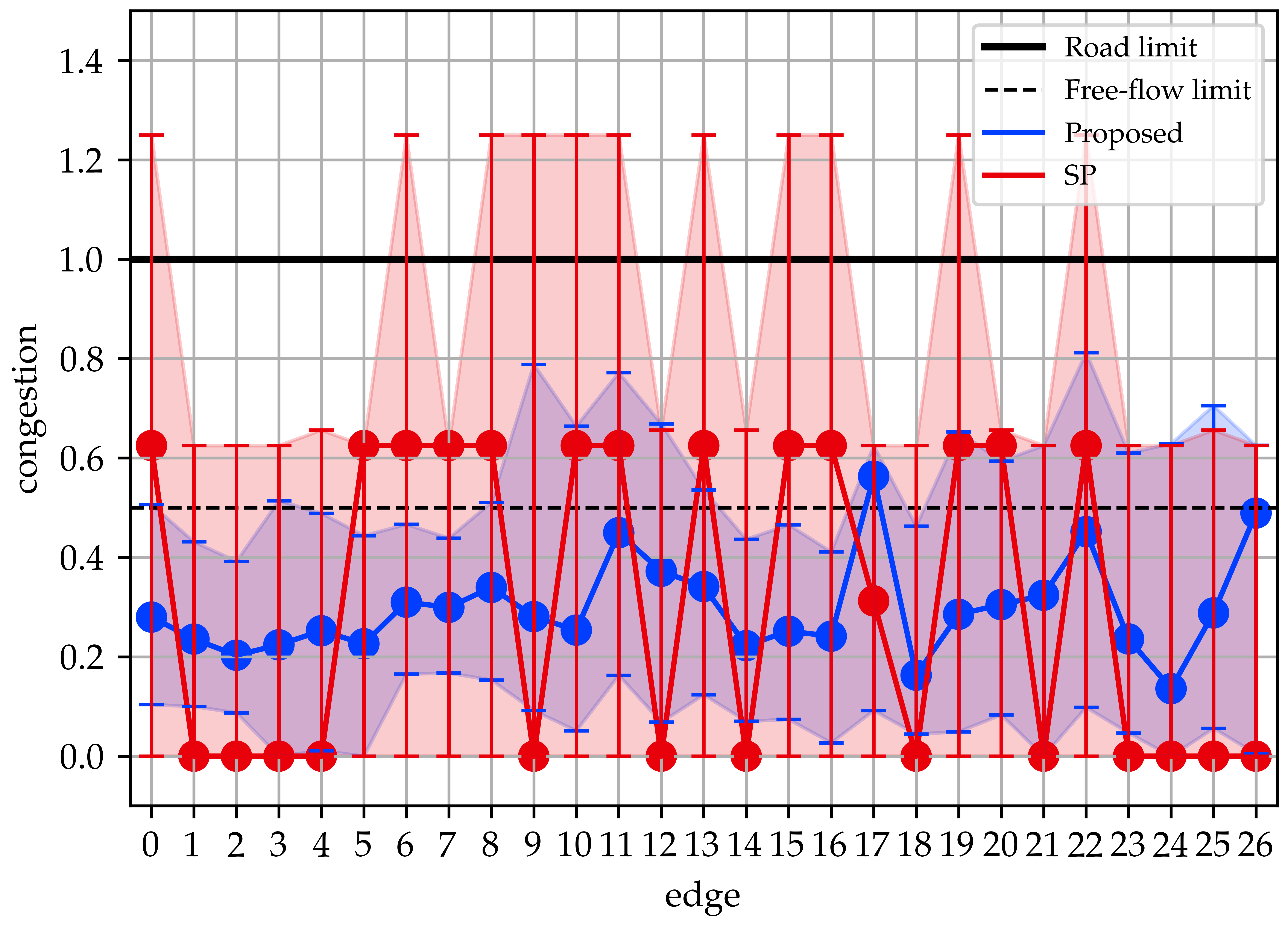}
	\caption{$\max_{t} \sigma_{(a,b)}^{t} / \overline{c}_{(a,b)}$, compared to the congestion obtained by the shortest path routing. The dotted line denotes  $c_{(a,b)} / \overline{c}_{(a,b)}$. The dots show the median values. The shaded area highlights the 95\% confidence interval. {We show in red the performance of the shortest path solution (SP).}	\label{fig:2_comparison_congestions} }
\end{figure}

\begin{figure}
	\centering
	\includegraphics[width=0.93\columnwidth]{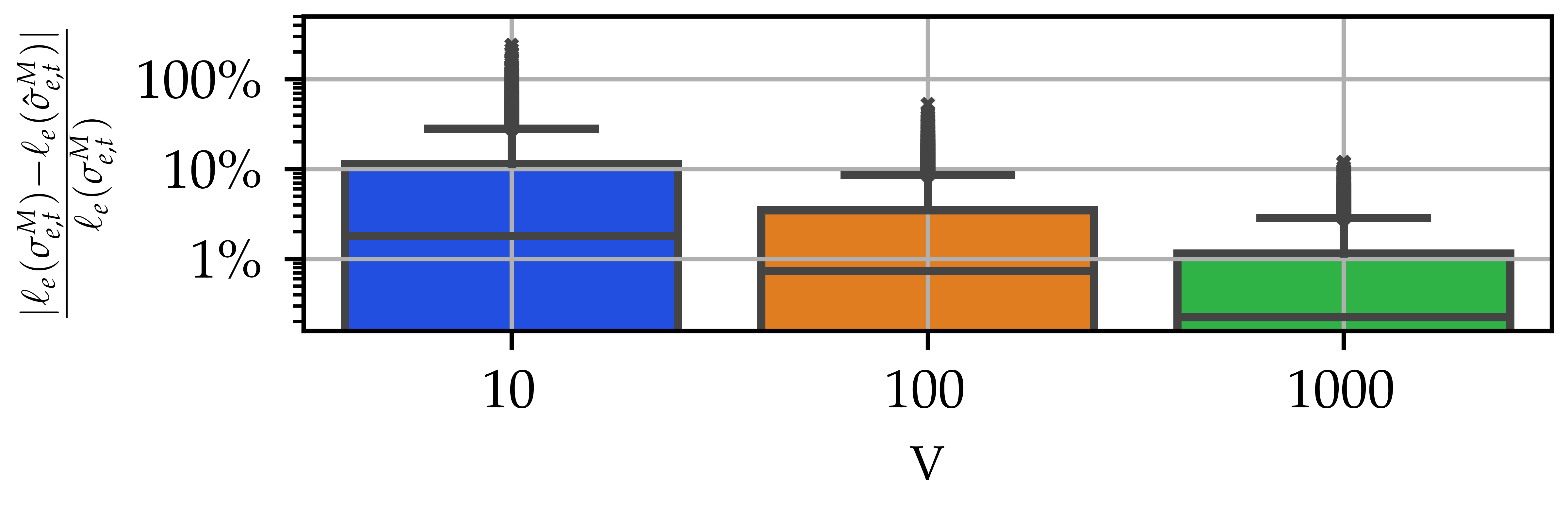}
	\caption{Difference between approximated and empirical travel time with respect to $V$, the number of vehicles per population. }
	\label{fig:4_comparison_expected_congestion}
\end{figure}


\begin{figure}
	\includegraphics[width=\columnwidth]{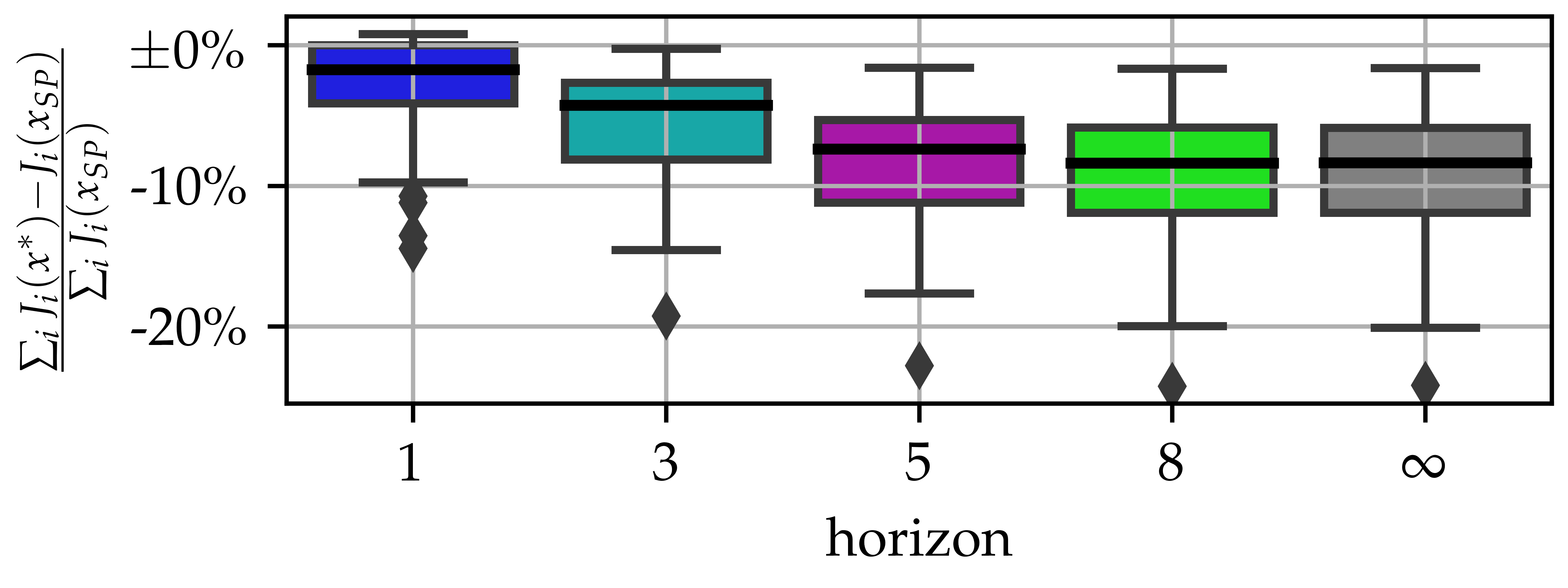}
	\caption{Comparison of the total cost incurred by the agents, with respect to the shortest path without traffic information.}
	\label{fig:2_multiperiod_advantage}
\end{figure}

\appendix
\begin{lemma} \label{le:eigenvalues_A}
	The only nonzero eigenvalues of a matrix
	\begin{equation}
		A(\by):= 2(\sigma +\zeta)\ones_N  \ones_N^{\top} + \tfrac{\xi}{N}(\by \ones_N^{\top} + \ones_N\by^{\top})
	\end{equation}
	where $\by\in\R^N_{\geq0}$, $\sigma:=\frac{1}{N}\sum_i [y]_i$, $\zeta\geq 0$, are $
	\lambda_- := \xi\sigma + \gamma_-$ and $
	\lambda_+ := \xi\sigma + \gamma_+$,
	where 
	{\small	\begin{align} \label{eq:definition_gamma}
			\begin{split}
				&\gamma_{\pm}\hspace{-1pt}:=\hspace{-1pt}N(\sigma\hspace{-1pt}+\hspace{-1pt}\zeta) \hspace{-1pt}\pm\hspace{-1pt} \sqrt{   N^2 (\sigma\hspace{-1pt}+\hspace{-1pt}\zeta)^2 \hspace{-1pt}+\hspace{-1pt} 2N\xi(\sigma\hspace{-1pt}+\hspace{-1pt}\zeta)\sigma \hspace{-1pt}+\hspace{-1pt} \tfrac{\xi^2\| \by \|^2}{N}   } .
			\end{split}
	\end{align}}
\end{lemma}
\emph{Proof (sketch):} 
As	$A(\by)$ is a sum of $3$ rank-1 matrices, it is at most rank $3$. We verify that $\lambda_{\pm}$ are eigenvalues with eigenvectors $\xi\by+\gamma_{\pm} \ones_N$. There is no third non-zero eigenvalue as $\mathrm{trace}(A(\by)) =\lambda_- + \lambda_+$.	\qedd
\subsection{Proofs of Section \ref{sec:traffic_GNE}}
\emph{Proof of Lemma \ref{le:M_reformulation}}:
We prove that \eqref{eq:nonlin_dyn} and  \eqref{eq:nonl_transform} hold true for the matrices computed as in \eqref{eq:inverse_transform}. We  note that, if $[\rho^i_t]_{\bar{a}}=0$ for some $\bar{a}\in\mc N$, by  \eqref{eq:M_constr_2} $\sum_{b: (\bar{a},b)\in\mc E} M^i_{t, (\bar{a}, b)} = 0$ and, from \eqref{eq:M_constr_3}, $M^i_{t, (\bar{a}, b)} = 0$ for all $b\in \mc N$. Substituting in \eqref{eq:inverse_transform}, we obtain \eqref{eq:nonl_transform}:
\begin{align*}
	\begin{split}
		[\Pi^i_t]_{(b,a)} [\rho^i_t]_a  \hspace{-1pt}&= \hspace{-1pt}\begin{cases}
			0 & \text{if}~ [\rho^i_t]_a =0 \\
			M^i_{t, (a,b)}    & \text{if}~ [\rho^i_t]_a \neq 0 
		\end{cases} = M^i_{t, (a,b)}
	\end{split}
\end{align*}
 By expanding the product $\Pi^i_t \rho^i_t$ and by substituting the latter and \eqref{eq:M_constr_1}, one obtains \eqref{eq:nonlin_dyn}. {Finally, we sum both sides of \eqref{eq:M_constr_1} and \eqref{eq:M_constr_2} for all $b\in\mc N$ and $a\in\mc N$, respectively, to obtain:
	$$ \textstyle\sum_{b\in\mc N}  [\rho_{t+1}^i]_b = \textstyle\sum_{(a,b)\in\mc E} M_{t,(a,b)}^i  = \sum_{a\in\mc N} [\rho_{t}^i]_a . $$
	By induction, $\rho_t^i\in \Delta^{|\mc N|}$  and $\col(M_{t,(a,b)}^i )_{(a,b)\in\mc E}\in \Delta^{|\mc E|}$. } \qedd

\par\emph{Proof of Proposition \ref{prop:justification_approx}}:
{As $\hat{\sigma}^{\text{M}}_{(a,b),t}$ is drawn from a Poisson binomial distribution scaled by $NV$, from \cite[Eq. 15]{wang_number_1993} and from $M^i_{(a,b),t}\in[0,1]$,}
$$\mathrm{Var}(\hat{\sigma}^{\text{M}}_{(a,b),t}) = \tfrac{1}{(NV)^2}\textstyle\sum_{i,v} (1-M^i_{(a,b),t}) M^i_{(a,b),t} \leq \tfrac{1}{4NV}.$$
By the Chebyschev's inequality, for any $\epsilon>0$ and $K_{\epsilon}=\frac{1}{\sqrt{\epsilon}}$,
\begin{equation*}
	\P\left[(\sigmaM_{(a,b),t} - \hat{\sigma}^{\text{M}}_{(a,b),t}) \geq \tfrac{K_{\epsilon}}{2\sqrt{NV}}  \right] \leq \epsilon.
\end{equation*}
The result then follows from \cite[Theorem 6.2.3]{wolter_introduction_2007} by using $r_n = \frac{1}{2\sqrt{NV}}$ and $\bs a= \sigma^{\text{M}}_{(a,b),t} $ (in the reference notation). \qedd

\subsection{Proofs of Section \ref{sec:GNE_seeking}}
\emph{Proof of Lemma \ref{le:cost_is_convex} (sketch)}:
Compute $J_{(a,b)}''(\cdot, \bs M^{-i}_{(a,b),t})$ for a generic  $(a,b), t, i$ and note that 
it is non-negative using  Assumption \ref{as:ell}. The result then follows by \cite[Prop. 8.14, 8,17]{bauschke_convex_2017}, and Assumption. \ref{as:local_cost}. \qedd 
\par \emph{Proof of Lemma \ref{le:monotonicity_decomposition}}:
Let us compute $F$:
\begin{align}
	\label{eq:pseudogradient_computed}
	\begin{split}
		&F(\bomega)= \mathrm{col}\left(\nabla f_i(\omega_i)\right)_{i\in\mc I} + \\
		& \mathrm{col}\left(\begin{bmatrix}
			\mathrm{col}\left( J_{(a,b)}'(\cdot, \bs M^{-i}_{t,(a,b)} )|_{M^i_{t, (a,b)}} \right)_{\substack{(a,b),t}} \\
			\bs 0_{|\mc N|(T+1)} 
		\end{bmatrix}\right)_{i\in\mc I},
	\end{split}
\end{align}
where the zero vector appears because the latency functions do not depend on $ (\rho^i_t)_{t, i}$. From Assumption \ref{as:local_cost} and \cite[Example 20.3]{bauschke_convex_2017}, $\nabla f_i $ is monotone for each $i$. Then, $\mathrm{col}(\nabla f_i)_i$ is  monotone by \cite[Prop. 20.23]{bauschke_convex_2017}.   Let us denote the second addend in \eqref{eq:pseudogradient_computed} as $T(\bomega)$. From \cite[Prop. 20.10]{bauschke_convex_2017}, $F$ is monotone if $T$ is monotone. Let us define the permutation matrix $P$ such that 
\begin{equation*}
	P\bomega = \begin{bmatrix}\mathrm{col}(\bs{M}_{t, (a,b)})_{\substack{(a,b)\in \mc E, t\in \mc T}} \\ \mathrm{col}(\bs{\rho}_t)_{\substack{ t\in\mc T^+ }}  \end{bmatrix}.
\end{equation*}
It holds, from the definition of $F_{(a,b),t}$,
\begin{equation}
	\label{eq:definition_PT}
	PT(\bomega) =  \begin{bmatrix}
		\mathrm{col}(F_{(a,b),t}(\bs{M}_{t, (a,b)}))_{\substack{(a,b)\in \mc E, t\in \mc T} }\\
		\bs 0_{N|\mc N|(T+1)}
	\end{bmatrix}. 
\end{equation}
As $PP^{\top}=I$, for all $\bomega, \by$:
\begin{align*}
	&\langle T(\bomega) - T(\by),\bomega - \by \rangle  = \langle PT(\bomega) - PT(\by), P\bomega - P\by\rangle =  \\ 
	&  \textstyle\sum_{(a,b), t} \langle F_{(a,b),t}|_{\bomega} - F_{(a,b),t}|_{\by}, \bs M_{t,(b,a)}|_{\bomega} - \bs M_{t,(b,a)}|_{\by} \rangle
\end{align*}
which is non negative if $F_{(a,b),t}$ is monotone $\forall (a,b),t$. \qedd
\par \emph{Proof of Lemma \ref{le:monotonicity_condition}}:
By \cite[Prop. 12.3]{rockafellar_variational_2009}, $T$ in \eqref{eq:operator_T} is monotone if $DT(\by)+DT(\by)^\top\succeq 0 ~\forall\by$. Denote $\sigma = \mathrm{avg}(\by)$.
\begin{align}\label{eq:jac_T}
	\begin{split}
		DT(\by) &=  \tfrac{1}{N} \ell'(\sigma) (I_N + \ones \ones^{\top}) + \tfrac{1}{N^2} \ell''(\sigma)  ( \by \ones^{\top}).
	\end{split}
\end{align}
As $\ell'(\sigma) = {k}(\sigma+\zeta)^{\xi} $,  $\ell''(\sigma) = {k\xi}(\sigma+\zeta)^{\xi-1}$, we  compute 
\begin{align}
	\label{eq:symmetric_jac}
	\begin{split}
		&DT(\by) + DT^{\top}(\by) = \tfrac{2k}{N}  (\sigma+\zeta)^{\xi}  I_N + \\
		&\tfrac{k}{N}  (\sigma+\zeta)^{\xi-1} (2(\sigma+\zeta)\ones\ones^{\top} + \tfrac{\xi}{N}( \by \ones^{\top} + \ones \by^{\top})).
	\end{split}
\end{align}
By Lemma \ref{le:eigenvalues_A}, $DT(\by) + DT^{\top}(\by)\succeq 0$ if 
\begin{equation}
	\label{eq:condition_posdef_1}
	\tfrac{2k}{N}(\sigma+\zeta)^\xi+ \tfrac{k}{N} (\sigma +\zeta)^{\xi-1}(\xi\sigma + \gamma_-(\by))\geq 0,
\end{equation}
where $\gamma_-$ is defined in \eqref{eq:definition_gamma}. Excluding the trivial case $\by=0, \zeta=0$, we divide by $\frac{k}{N}(\sigma+\zeta)^{\xi}$ to obtain
\begin{align}
	& \eqref{eq:condition_posdef_1}\Leftrightarrow {2} + \tfrac{\xi\sigma}{\sigma+\zeta} + \tfrac{\gamma_-(\by)}{\sigma+\zeta} \geq 0 \Leftrightarrow \nonumber \\
	& {2} + \tfrac{\xi\sigma}{\sigma+\zeta} + {N}  \geq \sqrt{ N^2 + 2 \tfrac{N\xi\sigma}{\sigma+\zeta} + \tfrac{\xi^2\| \by \|^2 }{N(\sigma+\zeta)^2} } \Leftrightarrow \nonumber \\
	&{4} + \tfrac{\xi^2\sigma^2}{(\sigma+\zeta)^2} + \cancel{{N^2}} + \tfrac{4\xi\sigma}{\sigma+\zeta} + 4{N} + \cancel{\tfrac{2N\xi\sigma}{\sigma+\zeta}} \geq \nonumber \\
	& \cancel{{N^2}} +  \cancel{\tfrac{2N\xi\sigma}{\sigma+\zeta}} + \tfrac{\xi^2\| \by \|^2 }{N(\sigma+\zeta)^2} \Leftrightarrow \nonumber \\
	& f(\by):=4(N+1) + \tfrac{\xi^2\sigma^2}{(\sigma+\zeta)^2} + \tfrac{4\xi\sigma}{\sigma+\zeta}  - \tfrac{\xi^2\| \by \|^2 }{N(\sigma+\zeta)^2}\geq 0. \label{eq:condition_posdef_2} 
\end{align}
We look for the minimum of the left hand side of the latter inequality.  Notice that $\nabla_{\by}\sigma = \frac{1}{N}\ones_N$. Then, 
\begin{align*}
	\nabla f(\by) &= \tfrac{2\xi^2}{N}\tfrac{ \sigma (\sigma+\zeta)^2 - \sigma^2(\sigma+\zeta)}{(\sigma+\zeta)^4}\ones_N - \tfrac{4\xi}{N}\tfrac{\zeta}{(\sigma+\zeta)^2} \ones_N \\
	&- \xi^2\tfrac{2N\by(\sigma+\zeta)^2 - 2(\sigma+\zeta)\|\by\|^2\ones_N}{N^2(\sigma+\zeta)^4}.
\end{align*}
Since $\nabla f(\by)$ contain either terms that multiply $\ones_N$ or $\by$, it must be $\by = \alpha \ones_N$ for some $\alpha \in (0,1]$ for $\by$ to be a stationary point. Therefore, the minimum of $f(\by)$ is either obtained for $\by=\alpha \ones_N$ or at an extreme point of $[0,1]^N$, that is, $\by=\sum_{i\in \mc Q} \boldsymbol{e}_i$, where $\bs e_i\in\R^N$ with only non-zero element $[\bs e_i]_i=1$ and $\mc Q \subset \{1,...., N\}$. Let us study the two cases separately:\\
\emph{Case $\by = \alpha\ones_N$:}
In this case, $\sigma = \alpha $ and $\|\by\|^2=\alpha^2 N$. We substitute these values in \eqref{eq:condition_posdef_2} to find
\begin{equation*}
	f(\by)=4(N+1) + \cancel{\tfrac{\xi^2\alpha^2}{(\alpha +\zeta)^2}} + \tfrac{4\xi\alpha}{\alpha+\zeta}  - \cancel{\tfrac{\xi^2\alpha^2N }{N(\alpha +\zeta)^2}} \geq 0.
\end{equation*}
\emph{Case $\by = \sum_{i\in \mc Q} \boldsymbol{e}_i$:} In this case, define $q:=|\mc Q|$, we compute $\sigma = \frac{q}{N}$ and $\| \by \|^2 = q$. We then substitute in \eqref{eq:condition_posdef_2} to find
\begin{equation*}
	f(\by)=4N+4 + \tfrac{\xi^2q^2}{(q+N\zeta)^2} + \tfrac{4\xi q}{q+N\zeta}  - \tfrac{\xi^2qN}{(q+N\zeta)^2} \geq 0.
\end{equation*}
A sufficient condition for the latter is that the first addend is greater than the negative one, which is true if
\begin{equation*}
	g(q):=4({q}+\zeta N)^2 - {q}\xi^2\geq 0.
\end{equation*}
Let us study the first derivative of $g$:
\begin{align*}
	\begin{split}
		g'(q) = 8\left( q+\zeta N \right) - \xi^2 \leq0 
		\Leftrightarrow q \leq \tfrac{\xi^2}{8} - \zeta N.
	\end{split}
\end{align*}
We conclude that $g(q)$ has the minimum in $q=1$ if $\zeta\geq \frac{\xi^2-8}{8N}$. We then note that $g(1)\geq 0$ if $\zeta\geq \frac{\xi-2}{2N}$. Therefore, $g(q)\geq 0 $ for all $q\in\{1,...,N\}$ if  \eqref{eq:monotonicity_condition} holds true, which in turns guarantees that \eqref{eq:condition_posdef_1} holds true for all $\by \in [0,1]^N$. \qedd

\par\emph{Proof of Proposition \ref{prop:main}}:
{
As $F_{(a,b),t}$ in \eqref{eq:def_F_ab} is in the form in \eqref{eq:operator_T}, $DF_{(a,b),t}$ can be computed as in \eqref{eq:jac_T}. Denote $\sigma=\mathrm{avg}(\by)$. For $\by\in [0,1]^N$, $\sigma \leq 1$  and $\|\bs y\|^2 \leq N$. 
Thus,
$$\| \by\ones_N^\top\|=\sqrt{\lambda_{\text{max}}(\by\ones_N^\top\ones_N\by)} = \sqrt{N\|\by\|^2}\leq N$$ 
From subadditivity, $\|\ones_N\ones_N^{\top}\|=N$ and the latter,
\begin{align*}
	&\max_{y\in[0,1]^N} \| DF_{(a,b),t}(\by) \| \leq \\
	&\max_{y\in[0,1]^N} \tfrac{1}{N}\ell_{(a,b)}'(\sigma)\| I + \ones\ones^\top\| + \tfrac{1}{N^2}\ell_{(a,b)}''(\sigma) \| \by\ones^\top\|\\
	& \leq \max_{y\in[0,1]^N} \tfrac{1}{N}\ell_{(a,b)}'(\sigma) (1+N) + \tfrac{1}{N^2}\ell_{(a,b)}''(\sigma)N \\
	&\overset{(\sigma \leq 1)}{\leq} \tfrac{k_{(a,b)}}{N} ((1 + \zeta_{(a,b)})^{\xi} + \xi(1 + \zeta_{(a,b)})^{\xi-1})= L_{(a,b)}.
\end{align*}
}
$F_{(a,b),t}$ is therefore $L_{(a,b)}$-Lipschitz continuous on $[0,1]^{N}$ following \cite[Thm. 9.2, 9.7]{rockafellar_variational_2009}. {As $\bs\Omega \subset [0,1]^{Nn_{\omega}} $, it can be shown that the second addend of \eqref{eq:pseudogradient_computed} is Lipschitz continuous with constant $\max_{(a,b)\in\mc E} L_{(a,b)}$ and thus $F$ is $L-$Lipschitz continuous. By  Lemma \ref{le:game_is_monotone} and \cite[Cor. 20.28]{bauschke_convex_2017}, $F$ is maximally monotone. Finally, by applying Lemma \ref{le:cost_is_convex}, all the assumptions of \cite[Thm. 1]{belgioioso_semi-decentralized_2023} are satisfied and the thesis follows.\qedd
\subsection{Proofs of Section \ref{sec:receding_horizon_GNE} }
\emph{Proof of Lemma \ref{le:congestion_must_affine}}:
 Denote by $F$ the pseudogradient of \eqref{eq:game2}. We compute
\begin{align*}
	&	DF(\bomega) =  \mathrm{diag}(\nabla^2 f_i(\omega_i))_{i\in \mc I} + \\
	&  \tfrac{1}{N}(I_N + \ones_N \ones_N^{\top} ) \otimes \begin{bmatrix}
		\displaystyle\diag (\ell_{(a,b)}')_{\substack{t \in \mc T, (a,b)\in\mc E}} & \bs 0 \\
		\bs 0 & \bs 0
	\end{bmatrix} + \\
	& \tfrac{1}{N^2}\col\left(  \ones_N^{\top} \otimes \begin{bmatrix}
		\displaystyle\diag(M^i_{t,(a,b)}\ell_{(a,b)}'')_{{t\in\mc T ,(a,b)  \in \mc E }}  & \bs 0 \\
		\bs 0 & \bs 0
	\end{bmatrix}  \right)_{i\in \mc I}.
\end{align*}
The arguments of $\col$ in the second addend are in general different for each $i\in \mc I$. Thus, $DF(\bomega)$ is  symmetric only if  $\ell_{(a,b)}''\equiv0$.  The thesis then follows from \cite[Theorem 4.5]{monderer_potential_1996}.  \qedd
\par\emph{Derivation of \eqref{eq:cost_rewritten}}:
\label{appx:derivation_rewritten_cost}
From Assm. \ref{as:linear_cost} and \ref{as:local_cost_separable} and $\Se^{(a,b)\top} (u_t^i+\ueq^i)=M^i_{t,(a,b)}$, we rewrite $J_i$ in \eqref{eq:obj_func} as:
\begin{align*}
	\begin{split} 
		& J_{i}(\omega_i)  = {f}_{i}^F (x_{T+1}^i)+  \textstyle\sum_{t}\left\{{f}^S_{i}(x_t^i, {u}_t^i)+ \textstyle\sum_{(a,b)} \hspace{-2pt} \left[ \left(\tau_{(a,b)} \right. \right.\right. \\
		&\left. \left.\left.+ \hspace{-1pt} \tfrac{k_{(a,b)}}{N} \textstyle\sum_j \left[ \Se^{(a,b) \top} (u_t^j +\ueq^j) \right]\right)\Se^{(a,b)\top} (u_t^i+\ueq^i)\right]\right\}\hspace{-2pt}.
	\end{split}
\end{align*}
Using the definitions of $C$ and $\bar{\tau}$ and rearranging,
\begin{align*}
	J_{i}(x_1^i, {\bu} ) =&f_{i}^F (x_{T+1}^i) + \textstyle\sum_{t} f^S_{i}(x^i_t, {u}^i_t) + \\
	& (\bar{\tau}^{\top} +\mathrm{avg}(\bs{u}_t + \bs{u}_{\text{eq}})^{\top}C) (u_t^i + \ueq^i).   
\end{align*}
From Assumption \ref{as:linear_cost} and the definition of $C$ and $\bar\tau$, $C\ueq^i= \bs 0$, $\bar{\tau}^{\top}\ueq^i= 0$ for any $i\in\mc I$, thus \eqref{eq:cost_rewritten} follows. \qedd
\par\emph{Proof of Lemma \ref{le:potential_of_MPC_game}}:
Let us denote
	${u}^i:=  \col(u_t^i)_{t\in\mc T}$; ${\bs u}=\col({u}^i)_{i\in\mc I}.$
We can rewrite  the agent cost in \eqref{eq:cost_rewritten} as 
\begin{align*}
	J_i(x_1^i, {\bu} )  &=  ( {f}_{i}^F  (x^i_{T+1})   + \textstyle\sum_{t\in\mc T}  f^{S}_{i}(x^i_t, u^i_t)  +  \bar{\tau}^{\top} u^i_t) \\
	& + \textstyle\sum_{j\in\mc I}\tfrac{1}{N}({u}^j)^\top  \hspace{-2pt}(I_T \otimes  C ) {u}^i  \end{align*}
The pseudo-gradient of \eqref{eq:MPC_game} reads then as  \cite[Eq. 32]{belgioioso_semi-decentralized_2023}
\begin{align*}
	F(\bx_1,{\bu}) \hspace{-1pt}&= \hspace{-1pt}\col(\nabla_{\bu_i}( {f}_{i}^F (x^i_{T+1}) \hspace{-1pt}+ \hspace{-1pt} \textstyle\sum_t f^{S}_{i}(x^i_t, u^i_t) \hspace{-1pt}+\hspace{-1pt} \bar{\tau}^{\top} u^i_t ))_{i\in\mc I}   \\ 
	&+ \tfrac{1}{N}(I_N + \ones_N\ones_N^{\top})\otimes (I_{T} \otimes C) {\bs u} . 
\end{align*}
{It can be verified by expanding the quadratic forms that 
	\begin{align*}
	&\textstyle\sum_{t=1}^N  \| \bu_t\|^2_{\bs 1_N\bs 1_N^{\top}\otimes C }= \|{\bu}\|^2_{\ones_N\ones_N^{\top} \otimes  (I_T \otimes C)} \\
	& 	\textstyle\sum_{t=1}^N \| \bu_t\|^2_{I_N \otimes C } = \| {\bs{u}}\|^2_{I_{TN} \otimes C }
	\end{align*}
}
By substituting the latter in the definition of $\potential$, one obtains
\begin{align*} \potential(\bx_1, \bu) & =  \textstyle\sum_i \left( {f}_{i}^F ({x}^i_{T+1})+  \textstyle\sum_{t}   f^{S}_{i}(x^i_t, u^i_t) + \bar{\tau}^{\top}  u^i_t \right)  +
	\\ & +\tfrac{1}{2N}\| {\bu}\|^2_{(I_N  + \ones_N\ones_N^{\top}) \otimes (I_T \otimes C)}  . 
\end{align*}
One can then compute $\nabla_{\bu}\potential$ to verify that it reads as $F$. \qedd
\par\emph{Proof of Theorem \ref{thm:main_receding_hor}}:
{Theorem \ref{thm:main_receding_hor} follows by verifying the hypothesis of \cite[Thm. 2.19]{rawlings_model_2017}. Namely, we prove a lower bound for the stage cost (Lemma \ref{le:stage_cost}) and that the terminal cost is a control Lyapunov function for the collective system $\bx_{t+1} = (I_N \otimes B)\bu_t$ (Lemma \ref{le:lyap_descent_1}). We first show some technical relations in Lemma \ref{le:relations_set_zeta}: }

\begin{lemma}
	\label{le:relations_set_zeta}
	The following  hold $\text{for all} ~(\bx, \bu)\in\mathbb{Z}, i \in \mc I$:
	\begin{subequations}
		\label{eq:relations_eequilibrium}
		\begin{align}
			\textstyle\sum_{\substack{a\neq b}}   (\Se^{(a,b)})^{\top}u_i  &= -(\Se^{(d_i,d_i)}) ^{\top} u_i; \label{eq:sum_inputs_err} \\
			  [x_i]_{d_i} &\geq (\Se^{(d_i,d_i)})^{\top} u_i  ~~ ; \label{eq:max_input_err} \\
			-[x_i]_{d_i} &\geq \max_{a\in\mc N} [x_i]_a . \label{eq:x_err_di_is_large} 
		\end{align}
	\end{subequations}
\end{lemma}
\emph{Proof.}
\emph{\eqref{eq:sum_inputs_err}}: From the definition of $\mathbb{Z}$, $Pu_i = x_i $. Substituting the definition of $P$ and summing each row,
\begin{equation}
	\label{eq:inputs_and_states_sum_to_0}
	\textstyle\sum_{a}\textstyle \sum_{b:(a,b)\in\mc E} \Se^{(a,b)\top} u_i  =\textstyle \sum_{a} [x_i]_a = 0
\end{equation}
where we used the definition of $\mathbb{X}_i$ and $\sum_{a\in\mc N} \rhoeq^i=1$.
Using the definition of $\mathbb{U}$, \eqref{eq:sum_inputs_err} follows by noting
$$ \textstyle\sum_{\substack{a\neq b}} \Se^{(a,b)\top}u_i {=} \sum_{(a,b)\in\mc E} \{\Se^{(a,b)\top} u_i\} - \Se^{(d_i,d_i)\top} u_i. $$
\emph{\eqref{eq:max_input_err}:}
From $u_i \in  \R_{\geq0}^{|\mc E|} - \{\ueq^i\} $  and $[\ueq^i]_a=0$ for each $a$ not associated to the edge $(d_i,d_i) $, it follows $(\Se^{(d_i,b)})^{\top}u_i\geq 0 $ for all $b\in\mc N\setminus \{d_i\}$. As $x_i = Pu_i$, {from the definition of $P$:}
\begin{align*}&
	[x_i]_{d_i} = \textstyle\sum_{b:(d_i,b)\in \mc E} (\Se^{(d_i,b)})^{\top} u_i \geq  (\Se^{(d_i,d_i)})^{\top} u_i.
\end{align*}
\emph{\eqref{eq:x_err_di_is_large}:}
From $x_i +  \rhoeq^i \in \Delta^{|\mc N|}$ {and  the definition of $\rhoeq^i$,} it follows $[x_i]_a\geq 0~ \forall~a\neq d_i$ and $[x_i]_{d_i}\leq 0$. Thus, 
\begin{align*}
	- [x_i]_{d_i} &\overset{\eqref{eq:inputs_and_states_sum_to_0}}{=} \textstyle\sum_{b\in\mc N \setminus\{d_i\}} [x_i]_b\geq  [x_i]_a  ~~ \forall~a\in\mc N. \hfill & \qedd
\end{align*}

\begin{lemma}
	\label{le:stage_cost}
	For all $(\bx,\bu)\in \mathbb Z$, the stage cost  in \eqref{eq:potential_of_game:stage} satisfies 
	\begin{equation}
		\label{eq:lower_bound_stage_cost}
		\potential^S(\bx, \bu) \geq \tfrac{\tau_{\text{min}} \|\bx\|}{N\nx}.
	\end{equation}
\end{lemma}
\begin{proof}
From ${C}\hspace{-1pt}\succeq\hspace{-1pt} 0$ and Assm. \ref{as:local_cost_separable}, $	\potential^S(\bx,\bu) \geq  \sum_i {\bar{\tau}}^{\top}u_i $. Thus,
\begin{align} 
	\begin{split}
		\label{eq:le:stage_cost:step_1}
		& 	\potential^S(\bx,\bu) \geq  \textstyle\sum_{i\in\mc I} \textstyle\sum_{\substack{(a,b)\in \mc E}} \tau_{(a,b)} (\Se^{(a,b)})^{\top} u_i   \\
		& \overset{\text{Ass. \ref{as:linear_cost}}} {\geq}  \tau_{\text{min}} \textstyle\sum_{i\in\mc I}\textstyle\sum_{(a,b)\in \mc E, a\neq b }   (\Se^{(a,b)})^{\top} u_i  .
	\end{split}
\end{align}
We then note
\begin{align*}
	\textstyle\sum_{a\neq b}  \hspace{-1pt} \Se^{(a,b)\top} \hspace{-1pt} u_i & \hspace{-2pt}\overset{\eqref{eq:sum_inputs_err}}{=} \hspace{-2pt} -\Se^{(d_i, d_i)\top} \hspace{-2pt} u_i \hspace{-3pt} \overset{\eqref{eq:max_input_err}}{\geq} \hspace{-3pt}  -[x^i]_{d_i}\hspace{-3pt}  =\hspace{-2pt}  |[x^i]_{d_i}|.
\end{align*}
Substituting in \eqref{eq:le:stage_cost:step_1},
\begin{equation}
	\label{eq:le:stage_cost:step_2}
	\potential^S(\bx,\bu) \geq \tau_{\text{min}}  \textstyle\sum_{i\in\mc I}|[x_i]_{d_i}| .
\end{equation}
From \eqref{eq:x_err_di_is_large}, $|[x_i]_{d_i}| = \| x_i \|_{\infty}$. Substituting in \eqref{eq:le:stage_cost:step_2},
\begin{equation*}
	\potential^S(\bx,\bu) \geq \tau_{\text{min}}  \textstyle\sum_{i\in\mc I} \| x_i \|_{\infty}.
\end{equation*}
We recall $\bx = \col_{i\in \mc I}(x_i)$, thus
\begin{equation*}
	\|\bx\|_{\infty} = \max_{ \substack{i,a}} | [x_i]_a | =  \max_{i\in\mc I} \|x_i\|_{\infty} \leq \textstyle\sum_{i\in \mc I}  \| x_i\|_{\infty}  .
\end{equation*}
We then obtain 
$\potential^S(\bx,\bu) \geq  \tau_{\text{min}} 	\|\bx\|_{\infty}.$
As for any $y\in\R^n$, $n\|y\|_{\infty} \geq \|y\|_2$ and since	$x\in \R^{N\nx}$, we obtain \eqref{eq:lower_bound_stage_cost}.
\end{proof}

\begin{lemma}
	\label{le:lyap_descent_1}
	Let $\potential^F$ be as in \eqref{eq:potential_of_game:final} and let Assumption \ref{as:cost_eval_wrt_shortest_path} and Equation \eqref{eq:condition_gamma_terminal} hold true. For all $\bx\in\mathbb{X}$, $(\bx,\barusp(x)) \in \mathbb{Z}$ and
	\begin{equation}\label{eq:le:lyap_descent:condition}
		\potential^F((I_N \otimes B)  \barusp(\bx)) - 	\potential^F(\bx)   \leq - \potential^S(\bx, \barusp(\bx)).
	\end{equation}
\end{lemma}
\begin{proof}
For compactness, we drop the dependencies of $\barusp$ on $\bx$ and we let $\bx^+ = ( I_N\otimes B) \barusp$. {Define the mapping from each node to all parent nodes along the known path $\NP{i}$:}
$$\PN{i}: \mc N \rightrightarrows \mc N , ~~ \PN{i}:b \mapsto \{ a: b=\NP{i}(a)\}. $$
Then, from the definition of $B$ and \eqref{eq:le:lyap_descent:input_choice},
\begin{align*}
\label{eq:le_lyap_descent_evolution_x}
\begin{split}
	[x_i^+]_ b &{=} \textstyle\sum_{a: (a,b)\in\mc E} (\Se^{(a,b)})^{\top} \usp{i}(x_i) \\ &{=} \textstyle\sum_{a\in\PN{i}(b)} [x_i]_{a} - \delta_{d_i}(a).
\end{split}
\end{align*}
From the definition of $\musp{i}$, 
\begin{equation}
\label{eq:le_lyap_descent:red_tau}
[\musp{i}]_{\NP{i}{(a)}} = [\musp{i}]_a - [\tausp{i}]_a . ~~ \text{for all}~ a\in\mc N.
\end{equation}
Then,	
\begin{align*}
&	(\musp{i})^{\top} x_i = \textstyle \sum_{a\in\mc N} [\musp{i}]_a [x_i]_a \\
&  =\textstyle \sum_{a\in\mc N} ([\musp{i}]_{\NP{i}{(a)}} + [\tausp{i}]_a) [x_i]_a  \\
& = (\tausp{i})^{\top} x_i + \textstyle \sum_{a\in \mc N}[\musp{i}]_{\NP{i}(a)}[x_i]_a   \\ 
& = (\tausp{i})^{\top} x_i + \textstyle \sum_{b\in \mc N}[\musp{i}]_{b} \textstyle\sum_{a\in\PN{i}(b)}[x_i]_a \\
& = (\tausp{i})^{\top} x_i + \textstyle \sum_{b\in \mc N}[\musp{i}]_{b} ([x_i^+]_ b  + \delta_{d_i}(b)).
\end{align*}
We note that $\NP{i}(d_i)=d_i$, thus $[\musp{i}]_{d_i}=0$ from Assumption \ref{as:linear_cost}, therefore $[\musp{i}]_{b}  \delta_{d_i}(b) = 0$ for all $b$. Thus, 
\begin{align*}
&	(\musp{i})^{\top} x_i = (\tausp{i})^{\top} x_i + (\musp{i})^{\top} x^+_i
\end{align*}
and, from Assumption \ref{as:cost_eval_wrt_shortest_path},
\begin{align}
	\begin{split}
	\label{eq:lyap_descent_1}
&\potential^F(\bx) \hspace{-1pt}- \hspace{-1pt}	\potential^F(\bx^+)  \hspace{-2pt} =  \hspace{-2pt} \textstyle\sum_{i} \hspace{-1pt} \sigma^F_i((\musp{i})^{\top} x_i) \hspace{-1pt} - \hspace{-1pt} \sigma_i^F((\musp{i})^{\top} x_i^+)  \\ 
& \geq m_F ((\bmusp)^{\top} \bx - (\bmusp)^{\top} \bx^+)=   m_F (\bstausp)^{\top} \bx. 
\end{split}
\end{align}
From the definition of $\bar{\tau}$ and from $\tau_{(d_i,d_i)}=0$, $\forall \bx \in \mathbb{X},$
\begin{align*}
&{\bar{\tau}}^{\top}  \usp{i}= \textstyle\sum_{(a,b)} \tau_{(a,b)} (\Se^{(a,b)})^{\top} \usp{i} \overset{\eqref{eq:le:lyap_descent:input_choice}}{=} \\ &\textstyle\sum_{a\in\mc N} \tau_{(a,\NP{i}(a))} ([x_i]_a - \delta_{d_i}(a)) {=} (\tausp{i})^{\top} x_i. 
\end{align*}
By Assumption \ref{as:cost_eval_wrt_shortest_path} and denoting $\bar{C} = (I_N + \bs 1\bs 1^{\top})\otimes C$,
\begin{align}
\label{eq:le:lyap_descent:step_2}
\begin{split}
	& \potential^{\text{S}}(\bx, \barusp) = \tfrac{ \|\barusp\|^2_{\bar{C}} }{2N}+ \textstyle\sum_{i}{f}^{\text{S}}_{i}(x_i, \usp{i}) + (\tausp{i})^{\top} x_i \\
	& \leq	\textstyle\sum_{i\in\mc I} (L_{S}+1) (\tausp{i})^{\top} x_i +  \tfrac{1}{2N} \|\barusp\|^2_{\bar{C}}  .
\end{split}	
\end{align}
From \eqref{eq:lyap_descent_1} and \eqref{eq:le:lyap_descent:step_2}, then \eqref{eq:le:lyap_descent:condition} holds if 
\begin{equation}
\label{eq:le:lyap_descent:step_3}
(m_F-1 - L_{\text{S}}) (\bstausp)^{\top} \bx \geq \tfrac{1}{2N} \|\barusp\|^2_{\bar{C}}.
\end{equation}

Let us find a lower bound for the LHS of \eqref{eq:le:lyap_descent:step_3}.
\begin{align}
\label{eq:le:lyap_descent_2:RHS_rewrite}
\begin{split}
	&(\bstausp) \hspace{-1pt}^{\top}  \hspace{-1pt}\bx \hspace{-2pt}= \hspace{-2pt} \textstyle\sum_{i, a}  \hspace{-1pt} [\tausp{i}]_a [x_i]_a \hspace{-2pt} \overset{\text{Ass.} \ref{as:linear_cost}}{=}  \hspace{-1pt}  \sum_{i} \hspace{-1pt}\sum_{a\neq d_i}  \hspace{-1pt}[\tausp{i}]_a [x_i]_a \\
	&\geq \tau_{\text{min}} \textstyle\sum_{i} \textstyle\sum_{a\neq d_i} [x_i]_a \overset{\eqref{eq:inputs_and_states_sum_to_0}}{=} \tau_{\text{min}} \textstyle\sum_{i}  (-[x_i]_{d_i}).
\end{split}
\end{align}
We now  rewrite the RHS of \eqref{eq:le:lyap_descent:step_3}:
\begin{align}
\label{eq:le:lyap_descent_2:quad_form_rewrite}
\begin{split}
	&\|\barusp\|^2_{\bar{C}}= \textstyle\sum_{i} \left( (\usp{i})^{\top} C \usp{i}  + \textstyle\sum_{j}\left( (\usp{j})^{\top} C \usp{i} \right)\right).
\end{split}
\end{align}
We then note that for all $i,j\in\mc N$, from the definition of $C$:
\begin{equation} 
\label{eq:le:lyap_descent_2:step_4}
(\usp{j})^{\top} C \usp{i} \hspace{-1pt}=\hspace{-1pt} \textstyle\sum_{(a,b)} k_{(a,b)} (\usp{j})^{\top} \Se^{(a,b)} (\Se^{(a,b)})^{\top} \usp{i}.
\end{equation}
From \eqref{eq:le:lyap_descent:input_choice}, $(\usp{j})^{\top} \Se^{(a,b)}\leq 1$ for all $(a,b)$ and $ (\Se^{(a,b)})^{\top} \usp{i}=0 $ if $b\neq\NP{i}(a)$. We continue from \eqref{eq:le:lyap_descent_2:step_4}:
\begin{align*}
&\leq  \textstyle\sum_{a\in\mc N} k_{(a,\NP{i}(a))} \Se^{(a,\NP{i}(a))} \usp{i}   \\
& = \textstyle\sum_{a \in\mc N} k_{(a,\NP{i}(a))} ([x_i]_a - \delta_{d_i}(a) )  \\
&= \textstyle\sum_{a \neq d_i} k_{(a,\NP{i}(a))} [x_i]_a \leq \bar{k} \textstyle\sum_{a\neq d_i} [x_i]_a  \overset{\eqref{eq:inputs_and_states_sum_to_0}:}{=} -\bar{k}  [x_i]_{d_i}
\end{align*}
where we noted $k_{(d_i,\NP{i}(d_i))}= k_{(d_i,d_i)} =0$ from Assumption \ref{as:linear_cost}. 
Substituting the latter in \eqref{eq:le:lyap_descent_2:quad_form_rewrite}, 
\begin{equation}\label{eq:le:lyap_descent_2:step_6}
\|\barusp\|^2_{\bar{C}}  \leq (N+1) \bar{k} \textstyle\sum_{i\in\mc I} (- [x_i]_{d_i}).
\end{equation}
From \eqref{eq:le:lyap_descent_2:step_6} and \eqref{eq:le:lyap_descent_2:RHS_rewrite}, \eqref{eq:le:lyap_descent:step_3} holds true under \eqref{eq:condition_gamma_terminal}.
\end{proof}
{We are now ready to present the proof of Theorem \ref{thm:main_receding_hor}:}
\begin{proof}
By \cite[Thm. 2]{scutari_potential_2006}, for any $\bx\in\mathbb{X}$, a solution of $\mathcal{G}(\bx)$ solves $\mathcal{O}(\bx)$. Then,  $\mathrm{col}(\kappa_i(\bx))_i$ is the first input of a sequence which solves \eqref{eq:MPC_opt_problem} with initial state $\bx$.
Problem \eqref{eq:MPC_opt_problem}  satisfies \cite[Assm. 2.2, 2.3]{rawlings_model_2017} under Assumptions \ref{as:local_cost} and \ref{as:linear_cost}. \cite[Assm. 2.14a]{rawlings_model_2017} follows from Lemma \ref{le:lyap_descent_1}. By Assumption \ref{as:local_cost}, $\potential^F$ is Lipschitz continuous. Thus, \cite[Assm. 2.14b]{rawlings_model_2017}  is satisfied by Lemma \ref{le:stage_cost}. $\mathbb{X}$ is control invariant for $\barusp(\cdot)$, as verified by computing $(I\otimes B)\barusp(\bx)$ for a generic $\bx\in\mathbb{X}$. \cite[Assm. 2.17]{rawlings_model_2017}  is then satisfied by applying \cite[Prop. 2.16]{rawlings_model_2017}. The thesis follows from \cite[Thm 2.19]{rawlings_model_2017}.
\end{proof}
\bibliography{bibliography_bibtex_new}

\begin{thebibliography}{10}

\bibitem{barth_traffic_2009}
M.~Barth and K.~Boriboonsomsin, ``Traffic congestion and greenhouse gases,''
  {\em Access Magazine}, vol.~1, no.~35, pp.~2--9, 2009.

\bibitem{european_commission_and_directorate-general_for_mobility_and_transport_sustainable_2019}
{\relax European Commission {and} Directorate-General for Mobility {and}
  Transport}, {\em Sustainable transport infrastructure charging and
  internalisation of transport externalities: executive summary}.
\newblock 2019.

\bibitem{jahn_optimal_1999}
O.~Jahn, R.~H. Möhring, and A.~S. Schulz, ``Optimal {Routing} of {Traffic}
  {Flows} with {Length} {Restrictions} in {Networks} with {Congestion},'' in
  {\em Operations {Research} {Proceedings}}, pp.~437--442, Springer, 1999.

\bibitem{jahn_system-optimal_2005}
O.~Jahn, R.~H. Möhring, A.~S. Schulz, and N.~E. Stier-Moses,
  ``System-{Optimal} {Routing} of {Traffic} {Flows} with {User} {Constraints}
  in {Networks} with {Congestion},'' {\em Operations Research}, vol.~53, no.~4,
  2005.

\bibitem{angelelli_system_2021}
E.~Angelelli, V.~Morandi, M.~Savelsbergh, and M.~Speranza, ``System optimal
  routing of traffic flows with user constraints using linear programming,''
  {\em European Journal of Operational Research}, vol.~293, pp.~863--879, Sept.
  2021.

\bibitem{wardrop_theoretical_1952}
J.~G. Wardrop, ``Some theoretical aspects of road traffic research,'' {\em
  Proceedings of the Institution of Civil Engineers}, no.~3, 1952.

\bibitem{roughgarden_how_2002-1}
T.~Roughgarden and E.~V. Tardos, ``How bad is selfish routing?,'' {\em Journal
  of the Association for Computing Machinery}, vol.~49, no.~2, 2002.

\bibitem{correa_selfish_2004}
R.~Correa and N.~E. Stier-Moses, ``Selfish {Routing} in {Capacitated}
  {Networks},'' {\em Mathematics of Operations Research}, vol.~29, no.~4, 2004.

\bibitem{dafermos_traffic_1972}
S.~C. Dafermos, ``The {Traffic} {Assignment} {Problem} for {Multiclass}-{User}
  {Transportation} {Networks},'' {\em Transportation Science}, vol.~6, no.~1,
  1972.

\bibitem{dafermos_traffic_1980}
S.~Dafermos, ``Traffic {Equilibrium} and {Variational} {Inequalities},'' {\em
  Transportation Science}, vol.~14, pp.~42--54, Feb. 1980.

\bibitem{larsson_augmented_1995}
T.~Larsson and M.~Patriksson, ``An augmented lagrangean dual algorithm for link
  capacity side constrained traffic assignment problems,'' {\em Transportation
  Research Part B: Methodological}, vol.~29, no.~6, 1995.

\bibitem{calderone_markov_2017}
D.~Calderone and S.~S. Sastry, ``Markov decision process routing games,'' {\em
  8th International Conference on Cyber-Physical Systems}, Apr. 2017.

\bibitem{li_tolling_2019}
S.~H.~Q. Li, Y.~Yu, D.~Calderone, L.~Ratliff, and B.~Acrkmese, ``Tolling for
  {Constraint} {Satisfaction} in {Markov} {Decision} {Process} {Congestion}
  {Games},'' in {\em American {Control} {Conference}}, IEEE, July 2019.

\bibitem{bakhshayesh_decentralized_2021}
B.~G. Bakhshayesh and H.~Kebriaei, ``Decentralized {Equilibrium} {Seeking} of
  {Joint} {Routing} and {Destination} {Planning} of {Electric} {Vehicles}: {A}
  {Constrained} {Aggregative} {Game} {Approach},'' {\em IEEE Transactions on
  Intelligent Transportation Systems}, 2021.

\bibitem{belgioioso_semi-decentralized_2023}
G.~Belgioioso and S.~Grammatico, ``Semi-{Decentralized} {Generalized} {Nash}
  {Equilibrium} {Seeking} in {Monotone} {Aggregative} {Games},'' {\em IEEE
  Transactions on Automatic Control}, vol.~68, Jan. 2023.

\bibitem{belgioioso_projected-gradient_2018}
G.~Belgioioso and S.~Grammatico, ``Projected-gradient algorithms for
  {Generalized} {Equilibrium} seeking in {Aggregative} {Games} are
  preconditioned {Forward}-{Backward} methods,'' in {\em {ECC}}, IEEE, 2018.

\bibitem{monderer_potential_1996}
D.~Monderer and L.~Shapley, ``Potential {Games},'' {\em Games and Economic
  Behavior}, vol.~14, pp.~124--143, 1996.

\bibitem{bauschke_convex_2017}
H.~H. Bauschke and P.~L. Combettes, {\em Convex {Analysis} and {Monotone}
  {Operator} {Theory} in {Hilbert} {Spaces}}.
\newblock {CMS} {Books} in {Mathematics}, Cham: Springer International
  Publishing, 2017.

\bibitem{benenati_tractable_2019}
E.~Benenati, M.~Colombino, and E.~Dall'Anese, ``A tractable formulation for
  multi-period linearized optimal power flow in presence of thermostatically
  controlled loads,'' {\em Conference on Decision and Control}, 2019.

\bibitem{u_s_b_of_public_roads_traffic_1964}
{\relax U. S. B. of Public Roads}, ``Traffic assignment manual for application
  with a large, high speed computer,'' {\em US Department of Commerce, Bureau
  of Public Roads}, 1964.

\bibitem{wang_number_1993}
Y.~H. Wang, ``On the number of successes in independent trials,'' {\em
  Statistica Sinica}, vol.~3, pp.~295--312, 1993.

\bibitem{wolter_introduction_2007}
K.~M. Wolter, {\em Introduction to variance estimation}.
\newblock Statistics for social and behavioral sciences, New York: Springer,
  2nd ed~ed., 2007.

\bibitem{facchinei_generalized_2010}
F.~Facchinei and C.~Kanzow, ``Generalized {Nash} {Equilibrium} {Problems},''
  {\em Annals of Operations Research}, vol.~175, pp.~177--211, Mar. 2010.

\bibitem{palomar_convex_2009}
D.~P. Palomar and Y.~C. Eldar, {\em Convex {Optimization} in {Signal}
  {Processing} and {Communications}}.
\newblock 2009.

\bibitem{rawlings_model_2017}
J.~B. Rawlings, D.~Q. Mayne, and M.~Diehl, {\em Model predictive control:
  theory, computation, and design}.
\newblock Madison, Wisconsin: Nob Hill Publishing, 2nd edition~ed., 2017.

\bibitem{scutari_potential_2006}
G.~Scutari, S.~Barbarossa, and D.~Palomar, ``Potential {Games}: {A} {Framework}
  for {Vector} {Power} {Control} {Problems} {With} {Coupled} {Constraints},''
  in {\em International {Conference} on {Acoustics} {Speed} and {Signal}
  {Processing} {Proceedings}}, IEEE, 2006.

\bibitem{rockafellar_variational_2009}
R.~Rockafellar, M.~Wets, and R.~Wets, {\em Variational {Analysis}}.
\newblock Grundlehren der mathematischen {Wissenschaften}, Springer Berlin
  Heidelberg, 2009.

\end{thebibliography}
\bibliographystyle{ieeetr}
\vspace{-15pt}
\begin{IEEEbiography}[{\includegraphics[width=1in,height
			=1.25in,clip, keepaspectratio]{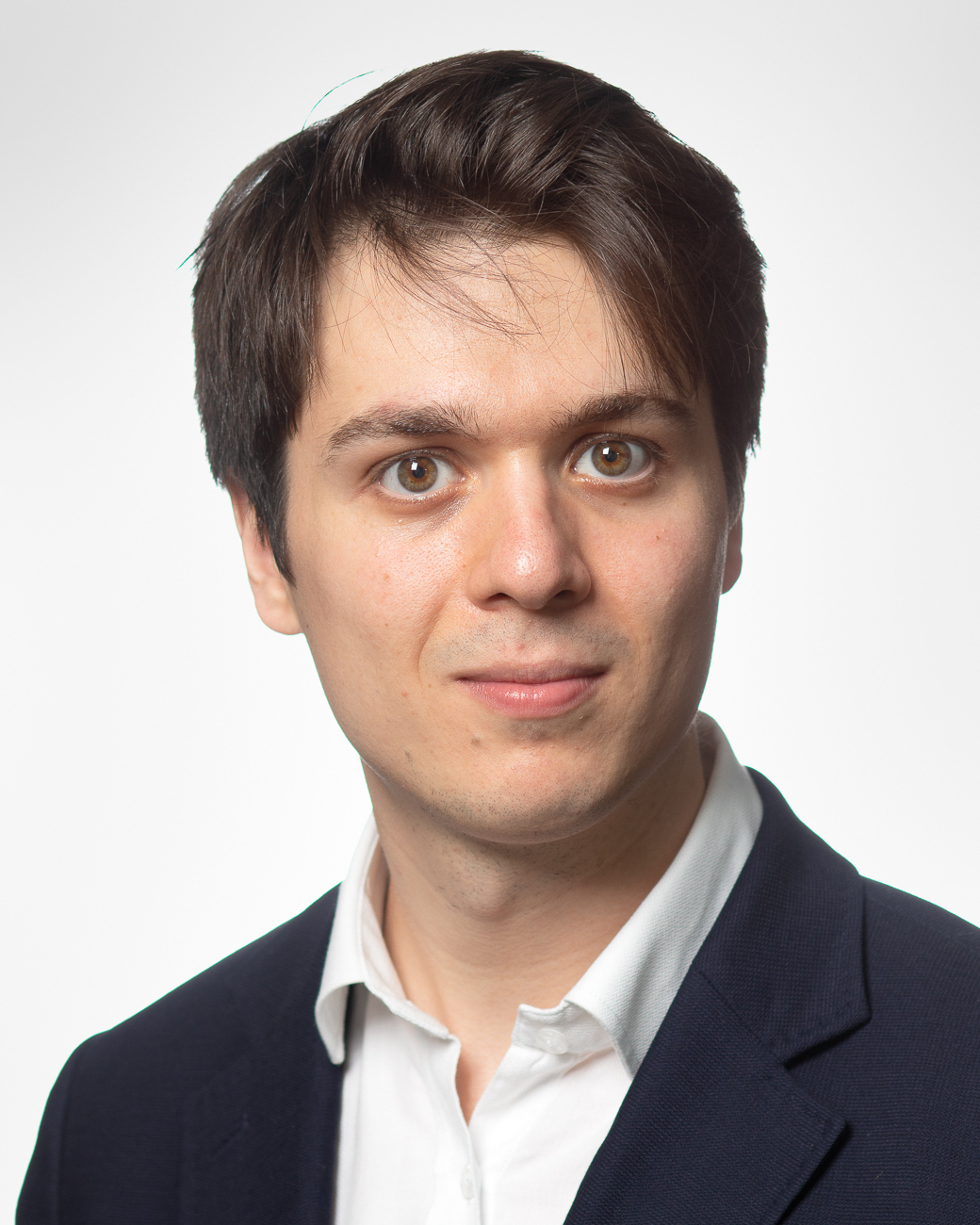}}]{Emilio Benenati} 
		is a doctoral student at the Delft Center for Systems and Control, TU Delft, The Netherlands. He received the Bachelor's degree from  the University of Catania, Italy in 2016 and his Master's degree from ETH Z\"urich, Switzerland, in 2019. He visited the University of Colorado Boulder, USA in 2018 and the University of California, Santa Barbara, USA, in 2024. In 2019-2020, he held a research position at the Italian Institute of Technology, Genova, Italy. 
\end{IEEEbiography} 
\vspace{-15pt}
\begin{IEEEbiography}[{\includegraphics[width=1in,height
			=1.25in,clip, keepaspectratio]{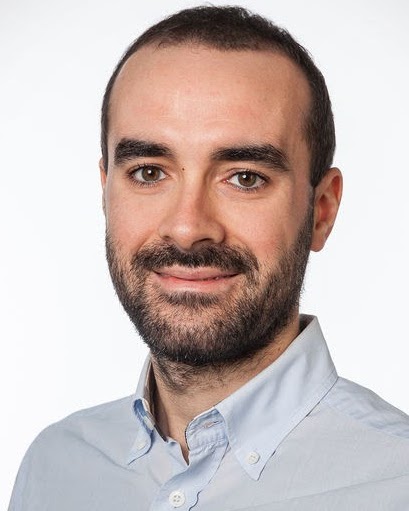}}]{Sergio Grammatico} 
		is an Associate Professor at the Delft Center for Systems and Control, TU Delft, The Netherlands. He received Bachelor, Master and PhD degrees from the University of Pisa, Italy, in 2008, 2009, 2013, respectively. In 2012-2017, he held research positions at UC Santa Barbara, USA, at ETH Zurich, Switzerland, and at TU Eindhoven, The Netherlands. Dr. Grammatico is a recipient of the 2021 Roberto Tempo Best Paper Award. He is an Associate Editor of the IEEE Transactions on Automatic Control and IFAC Automatica.
	\end{IEEEbiography}
\end{document}